\documentclass[12pt,letterpaper]{article}

\usepackage[left=1in,top=1in,right=1in,bottom=1.1in]{geometry}

\usepackage{graphicx}
\usepackage{booktabs}
\usepackage{multirow}
\usepackage{setspace}
\usepackage{url}
\usepackage[most]{tcolorbox}
\newtcolorbox{remarkbox}{
  colback=white, colframe=black,
  boxrule=0.5pt, arc=2pt,
  left=6pt, right=6pt, top=3pt, bottom=3pt,
  title=Remark
}
\usepackage{mathpazo}
\usepackage[T1]{fontenc}
\usepackage{color}
\definecolor{darkred}{RGB}{117,0,20}

\usepackage{titlesec}
\titleformat{\section}
  {\normalfont\fontsize{16}{0}\bfseries}{\thesection}{1em}{}
\titleformat{\subsection}
  {\normalfont\fontsize{14}{0}\bfseries}{\thesubsection}{1em}{}
\titleformat{\subsubsection}
  {\normalfont\fontsize{12}{0}\bfseries}{\thesubsubsection}{1em}{}
\titleformat{\paragraph}
  {\normalfont\fontsize{12}{0}\bfseries\itshape}{\theparagraph}{1em}{}
\titleformat{\subparagraph}
  {\normalfont\fontsize{12}{0}\itshape}{\thesubparagraph}{1em}{}
\makeatletter
\newcounter{subsubparagraph}[subparagraph]
\renewcommand\thesubsubparagraph{%
  \thesubparagraph.\@arabic\c@subsubparagraph}
\newcommand\subsubparagraph{%
  \@startsection{subsubparagraph}    
    {6}                              
    {\parindent}                     
    {12pt} 
    {6pt}                           
    {\normalfont\fontsize{12}{0}}}
\newcommand\l@subsubparagraph{\@dottedtocline{6}{10em}{5em}}
\newcommand{\subsubparagraphmark}[1]{}
\makeatother
\titlespacing*{\section}{0pt}{10pt}{4pt}
\titlespacing*{\subsection}{0pt}{6pt}{0pt}
\titlespacing*{\subsubsection}{0pt}{6pt}{1pt}
\titlespacing*{\paragraph}{0pt}{3pt}{3pt}
\titlespacing*{\subparagraph}{0pt}{3pt}{3pt}
\titlespacing*{\subsubparagraph}{0pt}{3pt}{3pt}

\usepackage[tableposition=top, figureposition=bottom, font=normalsize, labelfont=bf]{caption}

\usepackage{fancyhdr}
\usepackage[scaled]{helvet}

\makeatletter
\renewcommand{\maketitle}{\bgroup
   \begin{center}
   \vspace{15pt}
   \textbf{{\fontsize{19.8pt}{36}\selectfont \textsf{\@title}}}\\
   \vspace{15pt}
   {\fontsize{12pt}{0}\selectfont \@author}
   \end{center}
}
\makeatother

\usepackage{float}

\usepackage{enumitem}
\usepackage{eso-pic}

\fancypagestyle{firstpage}{
\fancyhf{}
\fancyhead[L]{%
  \includegraphics[height=0.7cm]{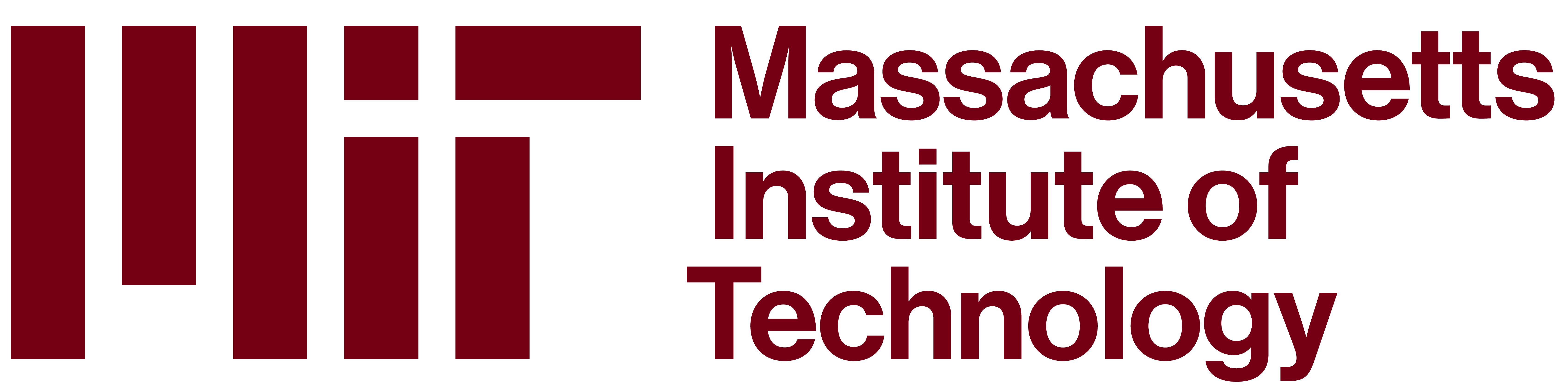}
}
\fancyhead[R]{}
\fancyfoot[C]{\thepage}

}

\fancypagestyle{fancy}{
  \fancyhf{}
  \fancyhead[C]{\footnotesize AI Fact-Checking in the Wild: A Field Evaluation of LLM-Written Community Notes on X}

  \fancyfoot[C]{\thepage}
}

\newenvironment{myquote}[1]%
  {\list{}{\leftmargin=#1\rightmargin=#1}\item[]}%
  {\endlist}

\renewenvironment{abstract}
{\vspace*{-.5in}\fontsize{12pt}{12}\begin{myquote}{0.3in}
\noindent \par{\bfseries \abstractname.}\par\noindent}
{\medskip\noindent
\end{myquote}
}
\usepackage{amsmath,amssymb,amsfonts}

\usepackage{tabularray}
\UseTblrLibrary{booktabs}
\usepackage{adjustbox}
\newcommand{\num}[1]{#1}
\usepackage{makecell}
\usepackage{subcaption}

\usepackage[authoryear,round]{natbib}
\usepackage[colorlinks=true,linkcolor=blue,citecolor=blue,urlcolor=blue]{hyperref}

\usepackage{listings}
\usepackage{xcolor}

\begin{document}

\title{AI Fact-Checking in the Wild: \\ A Field Evaluation of LLM-Written Community Notes on X}

\author{%
  Haiwen Li ~ ~ ~
  Michiel A. Bakker \\ [0.3em]
  Massachusetts Institute of Technology
}

\maketitle

\vspace{1.5em}

\begin{abstract}
Large language models (LLMs) show promising capabilities for fact-checking, yet prior work evaluates them only in controlled offline settings using benchmarks or crowdworker judgments. Success in real-world fact-checking depends also on how content is judged within a live platform environment. We present the first field evaluation of LLM fact-checking deployed on a live social media platform, testing performance directly through X Community Notes' ``AI writer'' feature over a three-month period. Our LLM writer, a multi-step pipeline that handles multimodal content, conducts web and platform-native search, and writes contextual notes, was deployed to write 1,614 notes on 1,597 tweets and compared against 1,332 human-written notes on the same tweets using 108,169 ratings from 42,521 raters. Direct comparison of note-level platform outcomes is complicated by differences in submission timing and exposure between LLM and human notes; we therefore pursue two analysis strategies: a rating-level analysis modeling individual rater evaluations, and a note-level analysis that relies on common raters who rated all notes on the same post. Rating-level analysis shows that LLM notes receive more positive ratings than human notes across raters with different political viewpoints, and note-level analysis shows LLM notes achieve significantly higher helpfulness scores among common raters. Rater-provided tags suggest that people consider LLM notes to use more neutral language and cite better sources. These findings provide field evidence that LLMs can contribute broadly helpful fact-checking notes at scale, while showing that their evaluation is shaped by platform dynamics absent from controlled offline settings.
\end{abstract}

\thispagestyle{firstpage}

\newpage
\section{Introduction}

Social media platforms increasingly rely on community-based mechanisms to add context to potentially misleading content. X Community Notes exemplifies this approach \citep{wojcik2022birdwatch}: users propose short, source-cited notes to add context to posts, and community ratings from a diverse group of raters determine whether a note is ``broadly helpful'' and should be shown publicly. This design creates a stringent alignment target that rewards neutral, evidence-grounded, cross-perspective explanations. Community-based fact-checking has also been examined to increase trust in fact-checking on social media and to reduce misinformation diffusion and engagement with misleading content \citep{drolsbach2024community, renault2024collaboratively, slaughter2025community}.

The core tasks of note authorship---researching relevant information, analyzing competing evidence, and writing clear factual explanations---align closely with capabilities where large language models have shown promise. Prior work finds that LLMs can generate fact-checking notes comparable to, and under certain conditions better than, human-written notes while requiring far less time and effort \citep{zhou2024correcting, de2025supernotes, singh2026gitsearch}. However, existing evaluations of LLM-based fact-checking remain offline: performance is measured against automated benchmarks, expert annotations, or crowdworker judgments in controlled settings \citep{zhou2024correcting, de2025supernotes, costabile2025assessing, singh2026gitsearch}. Crowdworker evaluations in particular are susceptible to demand effects where people evaluate content partly in response to what they infer the study expects \citep{mummolo2019demand, iarygina2025demand}. Hence, these evaluations fail to capture the on-platform dynamics such as timing, interface presentation, limited user attention, and subjective judging criteria that shape real-world impact. A critical gap therefore remains: we lack evidence on how LLM-generated fact-checks perform in situ, where their ultimate efficacy is determined not by gold-standard labels but on the dynamics of a live platform environment.

We present the first field evaluation of LLM fact-check writing deployed on-platform. X's recently released Community Notes ``AI writer'' API enables users to build AI writers that propose notes on potentially misleading posts flagged by human users.\footnote{\href{https://communitynotes.x.com/guide/en/api/overview}{https://communitynotes.x.com/guide/en/api/overview}} Notes generated by the system are clearly labeled as AI-created and are evaluated under the same crowd-sourced rating criteria as human-written notes. We developed and deployed our own LLM-based Community Notes writer on X through the ``AI writer'' API. This LLM writer conducts web search and platform-native search to collect information, assesses whether sufficient evidence exists to warrant a note, generates contextual notes, applies quality control filters, and publishes notes directly to the platform. The writer also handles multimodal posts that include images and videos. Over a three-month deployment period (November 1, 2025 to January 31, 2026), our LLM writer published 1,614 notes on 1,597 unique tweets, which we compared against 1,332 human-written notes on the same tweets, drawing on 108,169 ratings from 42,521 Community Notes raters. While there are multiple AI writers operating on the platform, we evaluate our own writer because its pipeline is fully documented and open-sourced, so our results can be interpreted in light of specific, reproducible design decisions rather than unobservable implementation. Notes from other AI writers are excluded from our analysis.

Our analysis reveals several key findings. First, direct comparison of note-level platform outcomes between LLM and human notes is complicated by systematic differences in submission timing created by platform design. Because notes submitted later receive fewer ratings and the platform's algorithm penalizes notes with fewer ratings, these timing differences can bias comparisons. We therefore pursue two analytical strategies. In a rating-level analysis, we show that LLM notes receive more positive evaluations from raters across the political spectrum, suggesting the potential of LLM-generated notes to achieve the cross-partisan consensus that Community Notes seeks to foster. In a note-level analysis that tries to mitigate the impact of different submission timing by relying on common raters who rated all notes on a given post, LLM notes achieve significantly higher helpfulness scores than human notes. Additional analysis on rater-provided tags also suggests that people consider the LLM notes to be less opinionated or biased and cite better sources.

Our contributions are threefold. Methodologically, we provide the first evaluation of LLM fact-checking in a real-world platform setting with organic user feedback, moving beyond laboratory studies, synthetic benchmarks, and proxy metrics to assess performance under authentic conditions, providing ecological validity. Practically, we open-source our writer implementation, including prompts and a multi-step writing pipeline we developed and refined iteratively over a month. Empirically, our findings demonstrate that LLMs can contribute high-quality fact-checking annotations at scale, achieving broader cross-ideological acceptance than human-written notes. We also show that real-world measures are more complicated than those measured in laboratory settings, and evaluation of AI notes in the wild requires special consideration of platform dynamics. 

\newpage
\section{Results}

\subsection{Platform Dynamics Complicate Note-Level Comparisons}

X Community Notes uses a matrix factorization--based bridging algorithm to aggregate ratings from individual users and evaluate note quality \citep{communitynotes_ranking}. Rather than relying on simple majority voting, the algorithm assigns higher helpfulness scores to notes that receive ``helpful'' ratings from raters who have historically disagreed in their past ratings. This design surfaces notes that are broadly acceptable across different viewpoints, and notes whose helpfulness scores exceed a platform-defined threshold receive the Currently Rated Helpful (CRH) status and become publicly visible on the platform.

Although note helpfulness scores and CRH status provide intuitive performance measures, comparing LLM- and human-written notes using these metrics is complicated by platform dynamics that are unrelated to intrinsic note quality. The Community Notes algorithm implicitly penalizes notes with fewer ratings and deflates their helpfulness scores,\footnote{Community Notes computes note helpfulness scores via matrix factorization with regularization that penalizes the note intercept term with a fixed shrinkage toward zero. This penalty is constant regardless of how many ratings a note receives, but the data signal that pushes the intercept away from zero scales with rating count. As a result, notes with fewer ratings have insufficient data to overcome the regularization pressure, leading to systematically deflated helpfulness scores.} yet the number of ratings a note receives depends heavily on when it is submitted. Earlier notes gain more exposure through longer visibility and more prominent placement, which in turn increases rating accumulation. During the deployment period of the study, LLM notes faced a systematic exposure disadvantage. The platform policy made posts available for LLM writers only after enough users have flagged the posts as potentially misleading, while human writers could propose notes whenever they notice misleading posts. Among the 814 tweets receiving both note types in our data, 66.0\% of human notes were created before the LLM notes. As a result, human notes accumulated substantially more ratings (mean $= 109.94$ vs.\ $59.50$; median $= 51$ vs.\ $22$; Mann--Whitney $U = 740{,}798.0$, $p < 0.001$). This temporal asymmetry implies that note-level metrics may partly reflect differences in exposure driven by submission timing rather than differences in note helpfulness alone. The order in which raters encountered multiple notes on the same post is also unknown. This is important because a rater may evaluate one note after learning information from another one, so the ratings are under the influence of cross-note learning.

To mitigate the influence of these confounds, we pursue two analysis strategies. First, we conduct analyses at the rating level, where each observation corresponds to an individual user's explicit evaluation of a note's helpfulness. These ratings directly capture users’ judgments of note helpfulness. We model individual ratings while accounting for raters' political ideology, aligning the analysis with the bridging principle underlying Community Notes. We treat this analysis as primary because it uses the full set of available ratings and focuses directly on user judgments of helpfulness rather than downstream platform outcomes affected by submission timing. Second, we perform a note-level analysis that recomputes note helpfulness scores using only ratings from common raters who saw and rated all notes on a given tweet, to reduce incomparability due to different submission timing. 

There is also the selection concern due to the different access to posts of LLM and human writers. This is challenging to solve through experimental design, because randomizing posts to human and LLM writers is not feasible on a live platform where human contributors organically choose which posts to annotate. The note-level analysis avoids post-level selection by comparing LLM- and human-written notes on the same posts, thereby holding post-level topic, factual difficulty, and visibility fixed. The rating-level analysis compares LLM notes with human notes on the posts for which the LLM writer submitted a note, preserving the platform environment in which note quality is evaluated, and we validate it with a robustness check that restricts the sample to posts that received both LLM and human notes (see \textit{SI}, Tables S8--S10).

\subsection{LLM Notes Receive More Positive Ratings Across the Ideological Spectrum}

Community Notes is designed around the principle of ``bridging'': a note is considered helpful if it is rated as such by raters with different viewpoints. The platform operationalizes this principle by estimating a numeric factor for each rater that captures systematic differences in rating behavior. In practice, this rater factor aligns closely with political ideology, with negative values corresponding to left-leaning rating patterns and positive values to right-leaning ones \citep{wojcik2022birdwatch}. Motivated by this design, our rating-level analysis examines how individual raters evaluate LLM notes versus human-written notes across the ideological spectrum. By leveraging the rater factor, we test whether LLM notes receive more positive ratings across raters with different inferred political viewpoints---the core criterion by which Community Notes defines helpfulness.

\begin{figure}
  \centering
  \includegraphics[width=\linewidth]{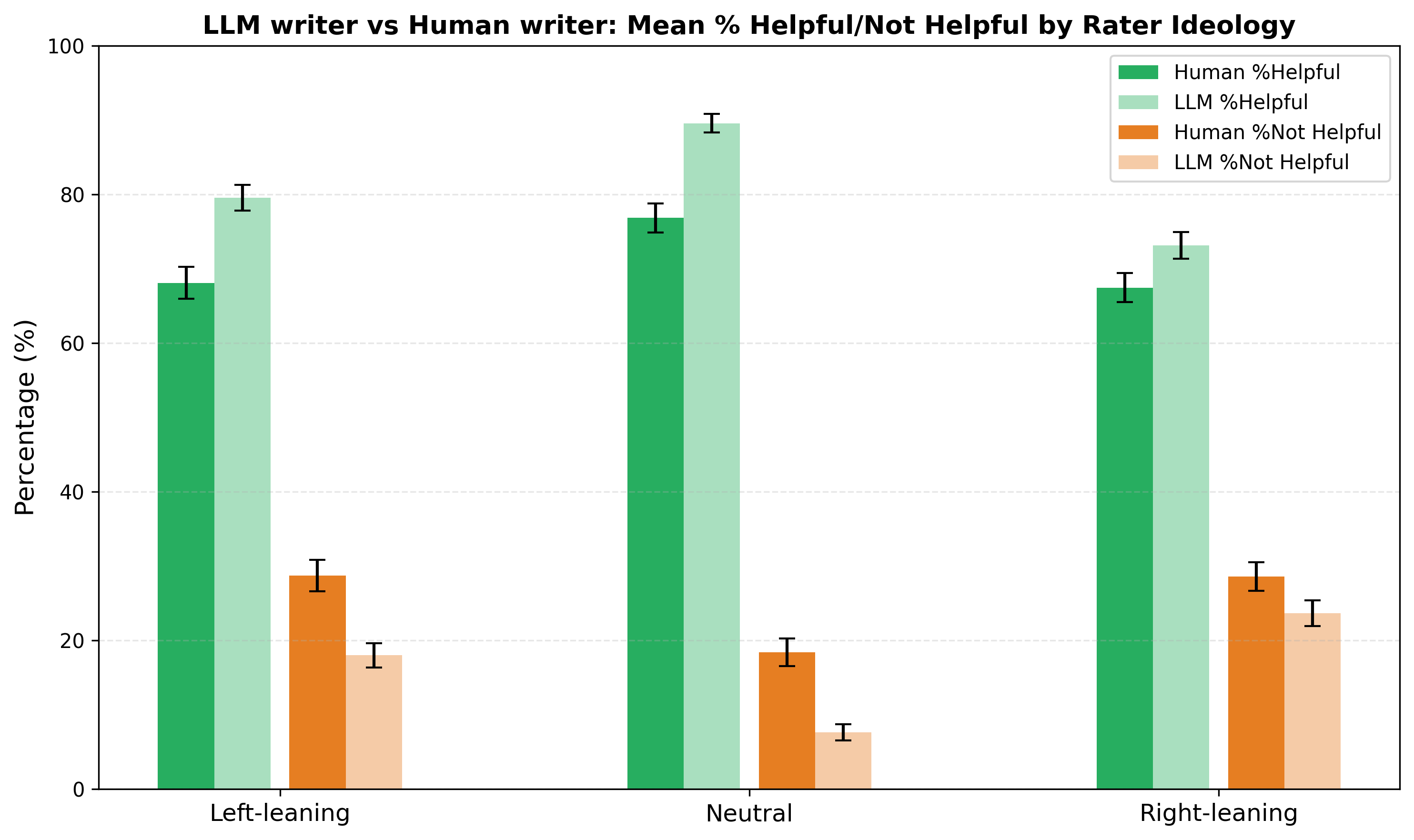}
  \caption{Mean \% helpful and \% unhelpful ratings per note for LLM and human notes, stratified by rater ideology group (left, neutral, right). Error bars show 95\% confidence intervals across notes.}
  \label{fig:1}
\end{figure}

To examine whether rating patterns varied across the political spectrum, we stratified raters into three ideology groups based on their rater factor following Community Notes convention: left (factor $< -0.15$), neutral ($-0.15 \leq$ factor $\leq 0.15$), and right (factor $> 0.15$). For each note and ideology group, we computed the percentage of ratings marked helpful and not helpful, then aggregated across notes to obtain mean percentages with 95\% confidence intervals. Results are presented in Figure~\ref{fig:1}. LLM notes have higher average\ \% helpful ratings and lower average\ \% unhelpful ratings than human notes across all rater ideology groups. The largest difference appears among neutral raters, followed by left-leaning raters, and then right-leaning raters.

To quantify differences in individual ratings between LLM and human notes, we estimated a linear mixed effects model to predict individual ratings from LLM writer authorship, rater political leaning, and their interactions. This analysis mirrors the ``bridging'' principle of Community Notes algorithm by modeling how the LLM advantage varies as a function of rater ideology, and the extremity of their ideology. Community Notes allows users to rate each note as Helpful, Somewhat Helpful, or Not Helpful; following the platform convention, we coded these responses numerically as 1.0, 0.5, and 0.0, respectively.

Specifically, we use the following primary specification:
\begin{multline}
  \text{rating\_score} \sim \text{AI} \times \text{coreRaterFactor1} \\
  + \text{AI} \times (\text{coreRaterFactor1})^{2} \\
  + (1 \mid \text{noteId}) + (1 \mid \text{raterId})
  \label{eq:main}
\end{multline}
where $\text{AI} = 1$ for LLM notes and $0$ for human notes, and $\text{coreRaterFactor1}$ is each rater's ideology factor, with negative values indicating left-leaning rating patterns and positive values indicating right-leaning patterns. The model includes both the linear and quadratic forms of the ideology factor, each interacted with LLM authorship, to allow the LLM effect to vary asymmetrically across the political spectrum, consistent with patterns observed in Fig.~\ref{fig:1}. The linear ideology term and its interaction with LLM capture directional differences in whether the LLM advantage is larger among left- or right-leaning raters. The quadratic term and its interaction capture extremity effects in whether ratings decline and the LLM advantage narrows as raters become more ideologically extreme in either direction. Note and rater random intercepts account for within-note correlation of ratings and between-rater heterogeneity in baseline rating tendencies. Table~\ref{tab:rating_reg_table} shows that LLM notes receive significantly more positive ratings at the center of the ideology spectrum (AI coefficient $= 0.104$, $p < 0.001$), given that human notes' average helpful ratings achieve around 78.5\% from centrist raters, this corresponds to approximately a 10-percentage-point increase. The negative quadratic interaction indicates that this advantage is largest among moderate raters and diminishes toward both ideological extremes, while the linear interaction indicates that the advantage decreases more steeply among right-leaning raters.

Because ratings may vary over the lifecycle of a note, we extended the rating-level specification to include the mean-centered log-transformed elapsed time between note creation and rating submission and its interaction with LLM authorship. The estimated LLM advantage remains similar and statistically significant after introducing the temporal structure (AI coefficient $=0.105$, $p<0.001$). Details are provided in \textit{SI}, Table S1.

\begin{table}[hbtp!]
\caption{Rating-Level linear mixed effects model results. To assess robustness, we estimate three additional specifications. Model 2 replaces note random intercepts with tweet random intercepts---a coarser grouping that absorbs tweet-level confounders while still identifying the AI advantage partly from within-tweet variation. Model 3 is an OLS regression with two-way cluster-robust standard errors clustered by note and rater. Model 4 replaces the continuous $\text{coreRaterFactor1}$ with categorical rater group (left-leaning, neutral, right-leaning; reference: neutral). The estimates of AI advantage stay consistent across all specifications.}
\centering
\footnotesize
\begin{adjustbox}{width=\linewidth}
\begin{tabular}{lcccc}
\toprule
& {Note + Rater RE}
& {Tweet + Rater RE}
& {OLS Clustered}
& \makecell{Note + Rater RE \\ (Group)} \\ \midrule
(Intercept)                                & \num{0.785}***  & \num{0.842}***  & \num{0.826}***  & \num{0.785}*** \\
                                           & (\num{0.006})   & (\num{0.005})   & (\num{0.010})   & (\num{0.007})  \\
AI                                         & \num{0.104}***  & \num{0.066}***  & \num{0.082}***  & \num{0.098}*** \\
                                           & (\num{0.009})   & (\num{0.004})   & (\num{0.012})   & (\num{0.010})  \\
coreRaterFactor1                           & \num{-0.015}*** & \num{-0.007}    & \num{-0.005}    &                \\
                                           & (\num{0.004})   & (\num{0.004})   & (\num{0.032})   &                \\
coreRaterFactor1\textsuperscript{2}        & \num{-0.175}*** & \num{-0.195}*** & \num{-0.291}*** &                \\
                                           & (\num{0.010})   & (\num{0.010})   & (\num{0.032})   &                \\
AI:coreRaterFactor1                        & \num{-0.087}*** & \num{-0.112}*** & \num{-0.120}*** &                \\
                                           & (\num{0.005})   & (\num{0.005})   & (\num{0.046})   &                \\
AI:coreRaterFactor1\textsuperscript{2}     & \num{-0.190}*** & \num{-0.169}*** & \num{-0.181}*** &                \\
                                           & (\num{0.013})   & (\num{0.013})   & (\num{0.048})   &                \\
left-leaning rater                         &                 &                 &                 & \num{-0.031}*** \\
                                           &                 &                 &                 & (\num{0.005})   \\
right-leaning rater                        &                 &                 &                 & \num{-0.059}*** \\
                                           &                 &                 &                 & (\num{0.005})   \\
AI:left-leaning rater                      &                 &                 &                 & \num{-0.003}    \\
                                           &                 &                 &                 & (\num{0.007})   \\
AI:right-leaning rater                     &                 &                 &                 & \num{-0.086}*** \\
                                           &                 &                 &                 & (\num{0.007})   \\
SD (Intercept raterParticipantId)          & \num{0.160}     & \num{0.160}     &                 & \num{0.164}     \\
SD (Observations)                          & \num{0.304}     & \num{0.335}     &                 & \num{0.304}     \\
SD (Intercept noteId)                      & \num{0.196}     &                 &                 & \num{0.197}     \\
SD (Intercept tweetId)                     &                 & \num{0.142}     &                 &                 \\
Num.Obs.                                   & \num{108169}    & \num{108169}    & \num{108169}    & \num{108169}    \\
\bottomrule
\end{tabular}
\end{adjustbox}
\par\vspace{4pt}
{\footnotesize\noindent * $p < 0.05$, ** $p < 0.01$, *** $p < 0.001$. Standard errors in parentheses. Models 1--2: model-based SEs from mixed model. Model 3: two-way clustered CR2 SEs by note and rater. Model 4: mixed model with categorical rater group. Reference category is neutral.}
\label{tab:rating_reg_table}
\end{table}

\begin{figure}
  \centering
  \includegraphics[width=\linewidth]{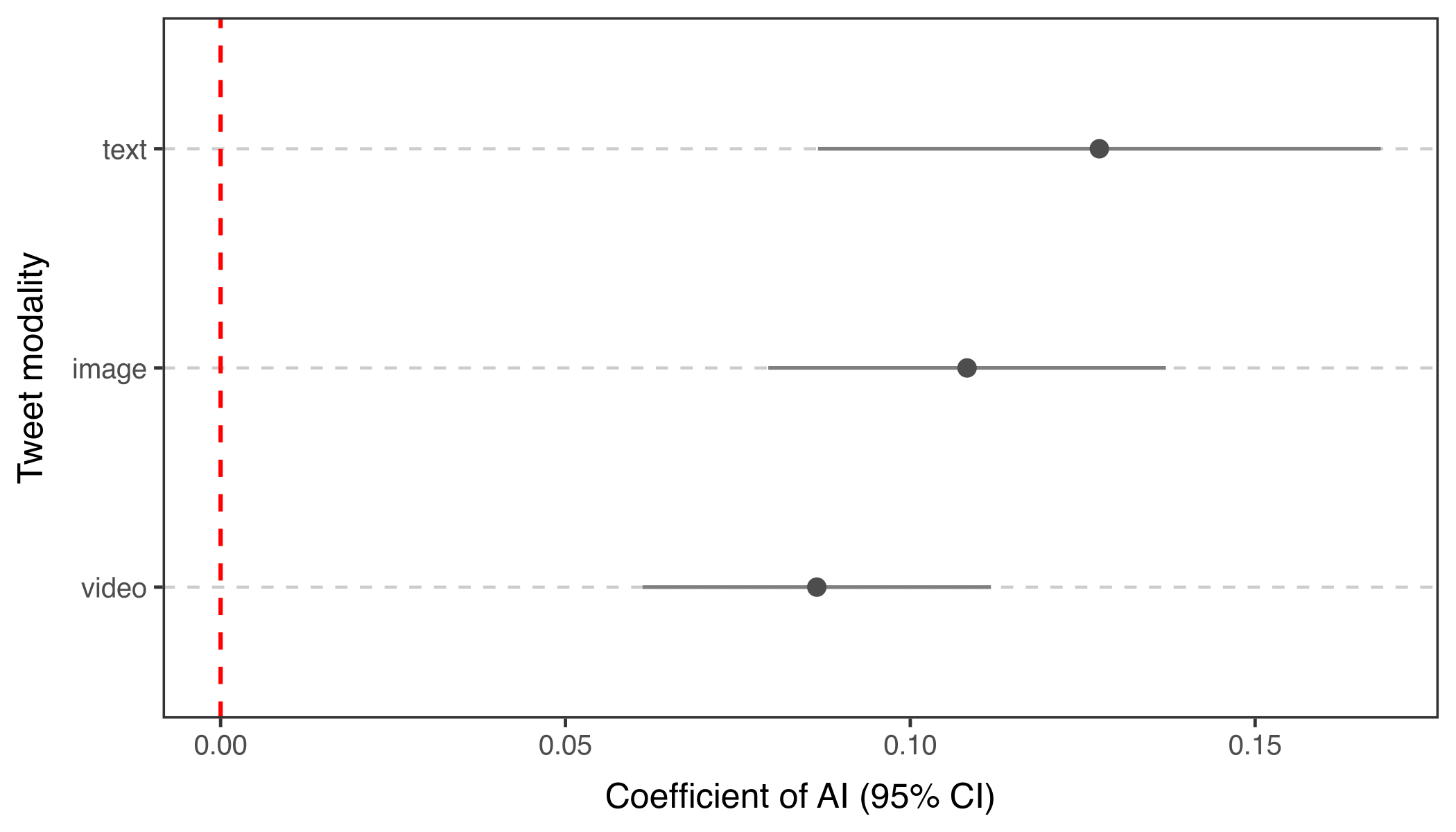}
  \caption{LLM vs.\ human note rating advantage (AI main-effect coefficient with 95\% CI) by tweet modality (text-only, image, video).}
  \label{fig:2}
\end{figure}

\begin{figure}
  \centering
  \includegraphics[width=\linewidth]{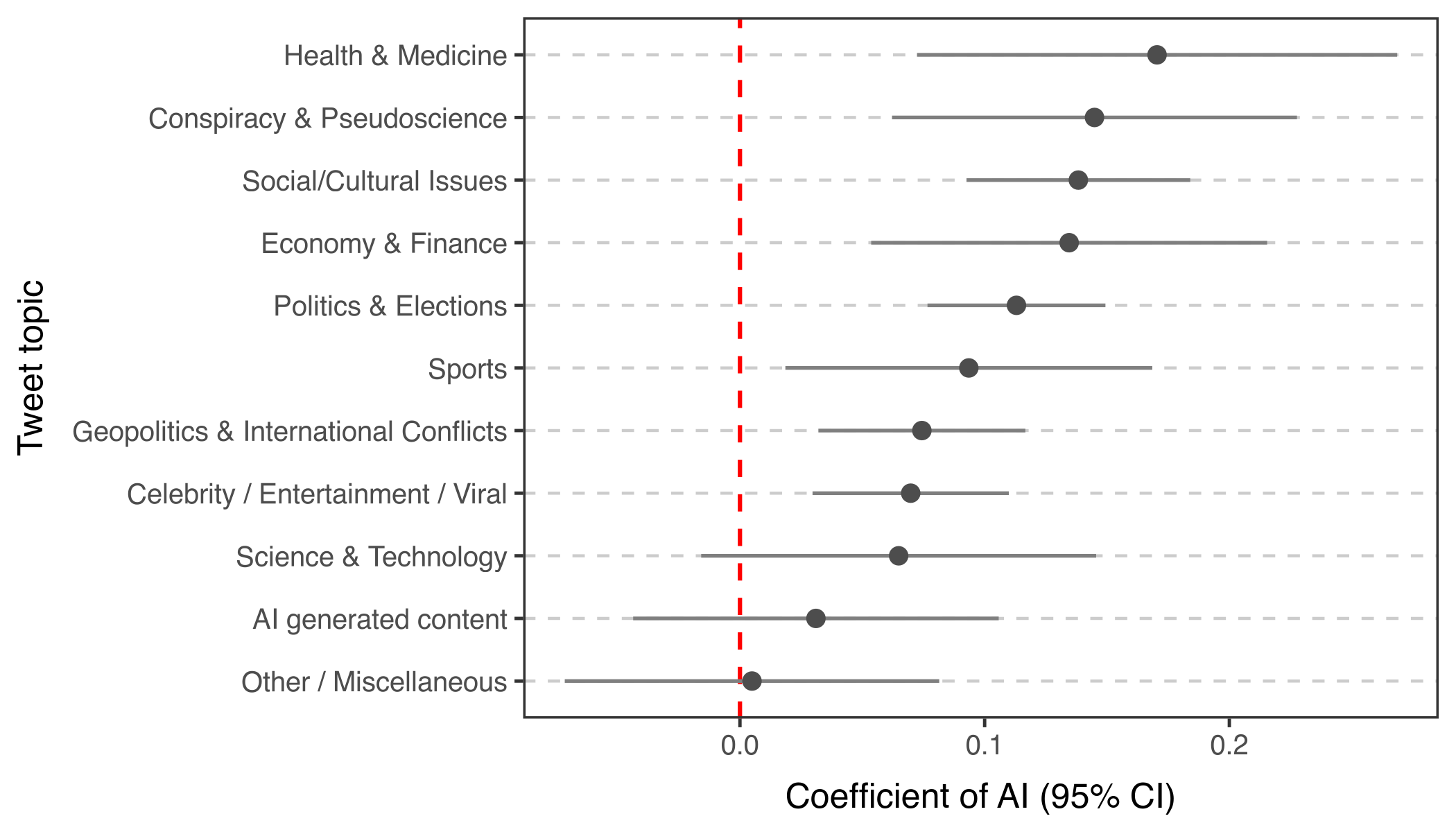}
  \caption{LLM vs.\ human note rating advantage (AI main-effect coefficient with 95\% CI) by tweet topic category.}
  \label{fig:3}
\end{figure}

To examine how the LLM advantage in helpfulness ratings varies across tweet types, we conducted exploratory subgroup analyses by tweet modality and topic. We used an LLM to assign a topic to each tweet based on its content (see \textit{SI} for the prompt). We subset notes by each category and re-estimated our main rating-level specification within each subset. Figs.~\ref{fig:2} and~\ref{fig:3} present the coefficient on the LLM main effect with 95\% confidence intervals. The LLM advantage is consistently positive across all three modality types but varies in magnitude: the point estimate is largest for text-only posts, followed by posts with images and videos. Topic-level heterogeneity is more pronounced. LLM notes receive better ratings on posts about health and medicine, and conspiracy theories and pseudoscience claims. Notably, LLM notes show minimal or non-significant advantages on posts about AI-generated content. These patterns suggest that the LLM writer is more effective at fact-checking claims in domains with well-established authoritative sources and clear factual grounding.

\newpage
\subsection{Common-Rater Note-Level Analysis}

The rating-level analysis demonstrates that LLM notes receive more positive individual evaluations across the ideological spectrum. We complement this with a note-level analysis to examine whether this advantage translates into better note-level outcomes. In the full sample, direct platform outcomes show mixed signals. LLM notes accumulate fewer ratings (median 22 vs. 51)\footnote{However, we find no evidence that LLM notes accumulated fewer ratings when LLM and human notes were written at approximately the same time, see \textit{SI}, Table S7.} and have a lower CRH rate (13.07\% vs.\ 18.02\%), while having similar note helpfulness scores (mean: 0.25 vs.\ 0.24) and a lower rate of reaching Currently Rated Not Helpful status (1.12\% vs.\ 4.13\%). However, as discussed above, direct comparison of platform note-level outcomes is confounded by differential exposure, because LLM notes are systematically submitted later and consequently accumulate fewer ratings. To address this, we construct a subset of ratings by retaining only ratings from raters who saw and rated every note on that tweet, ensuring that each note within a tweet received evaluation from the same set of raters. We then recompute note helpfulness scores by applying the Community Notes scoring algorithm to these common-rater ratings alone. This yields a subset of 13,721 ratings across 1,674 notes on 663 tweets. We compare both the characteristics of these common raters and the topics of the included posts with those in the full sample, and find the common raters do not differ substantially from the broader rater population in their overall tendency to rate notes as helpful or in their political leaning, and the topic distribution of posts included in the common-rater analysis does not differ significantly from that of the full set of posts (see \textit{SI}, Fig.~S1, for details).

We then estimate the differences in re-computed note-level outcomes with a linear mixed effects model:
\begin{equation}
    \text{outcome} \sim \text{AI} + (1|\text{tweetId})
    \label{eq:2}
\end{equation}
where $\text{AI} = 1$ for LLM-written notes and $0$ for human-written notes, and tweet random intercepts account for within-tweet dependence and absorb tweet-level confounders such as topic difficulty and post virality. We use mixed models rather than standard parametric tests because notes targeting the same tweet are not independent, violating the independence assumption. We use this model to analyze three note-level outcomes: note helpfulness score, CRH status, and CRNH status, and report Benjamini--Hochberg adjusted p-values.

Within the common-rater subset, LLM notes achieve significantly higher helpfulness scores than human notes (mean: 0.21 vs.\ 0.18; AI coefficient $= 0.019$, $z = 2.944$, adj. $p = 0.010$), confirming that the LLM quality advantage identified in the rating-level analysis carries through to the note-level metric once differences in the set of raters evaluating each note are removed. We also derive final rating status from the recomputed helpfulness scores: a higher proportion of LLM notes reach CRH status (2.40\% compared with 1.89\% for human notes) and a lower proportion reach CRNH status (0.90\% compared with 1.39\% for human notes), though these differences are not statistically significant. Because restricting to common raters substantially reduces the number of ratings per note, both the note helpfulness scores and the proportion of notes reaching CRH status are lower than those observed in the full sample. We further report note-level robustness checks in \textit{SI}, Tables S6 and S7, including subsetting the sample to notes with different minimum rating thresholds to mitigate the algorithm's penalty on low-rating-count notes and using creation-time--matched analyses to control for differences in exposure.

\subsection{Note Characteristics and Source Use}

LLM and human notes differed in both writing and source-citation behavior. LLM notes were much longer than human notes (mean 35.8 words vs.\ 26.9 words; $t = 22.190$, $p < 0.001$) and cited more URLs (mean: 1.51 vs.\ 1.23 URLs; $t = 8.67$, $p < 0.001$). To further characterize differences in note content, we measure readability, valence, and toxicity after removing URLs from the note text \citep{hutto2014vader, Detoxify}. LLM-written notes are harder to read (Flesch-Kincaid grade level: 9.96 versus 7.40; $t = 18.16$, $p < 0.001$), more negative (VADER valence score: $-0.279$ versus $-0.086$; $t = -11.83$, $p < 0.001$), and have lower predicted toxicity (0.007 versus 0.020; $t = -4.68$, $p < 0.001$; predicted toxicity was very low overall).

LLM and human notes also show different sourcing patterns (Tables~\ref{tab:top10_llm_domains} and~\ref{tab:top10_human_domains}). LLM notes most frequently referenced mainstream news outlets and social media platforms, including reuters.com (7.7\% of LLM notes), en.wikipedia.org (7.2\%), instagram.com (5.8\%), youtube.com (5.3\%), x.com (4.2\%), bbc.com (4.2\%), snopes.com (3.9\%), cnn.com (2.4\%), facebook.com (2.0\%), and yahoo.com (2.0\%). In contrast, human notes most commonly cited x.com (18.7\%), nearly four times the rate observed in LLM notes (4.2\%). Wikipedia appeared at similar rates in both groups (7.6\% human vs.\ 7.2\% LLM), but other mainstream news sources appeared less frequently in human notes: reuters.com (1.6\% vs.\ 7.7\%), bbc.com (1.1\% vs.\ 4.2\%), and snopes.com (0.7\% vs.\ 3.9\%). This pattern suggests that our LLM writer relies more heavily on traditional authoritative sources, whereas human writers more frequently reference platform-native content and social media posts. We note that this pattern characterizes our specific implementation.

To evaluate the quality of sources cited in LLM- and human-written notes, we compare the factual reporting and credibility of their cited domains. We use source ratings from Media Bias/Fact Check (MBFC), matching cited URLs to MBFC domains after normalizing hostnames and allowing parent-domain matches for subdomains. MBFC provides source-level measures of factual reporting and credibility; we code factual reporting on a 0--5 scale from Very Low to Very High and credibility on a 0--2 scale from Low to High. MBFC matched 1,411 of the 2,436 URLs cited by LLM notes (57.9\%) and 629 of the 1,637 URLs cited by human notes (38.4\%). At the note level, 1,091 of 1,614 LLM notes had at least one matched URL (67.6\%), compared with 468 of 1,332 human notes (35.1\%). The largest unmatched domains are social media platforms, including x.com, youtube.com, instagram.com, reddit.com, and tiktok.com, accounted for 29.0\% of unmatched URLs from LLM notes and 41.6\% of unmatched URLs from human notes. 

\begin{figure}[H]
  \centering
  \includegraphics[width=\linewidth]{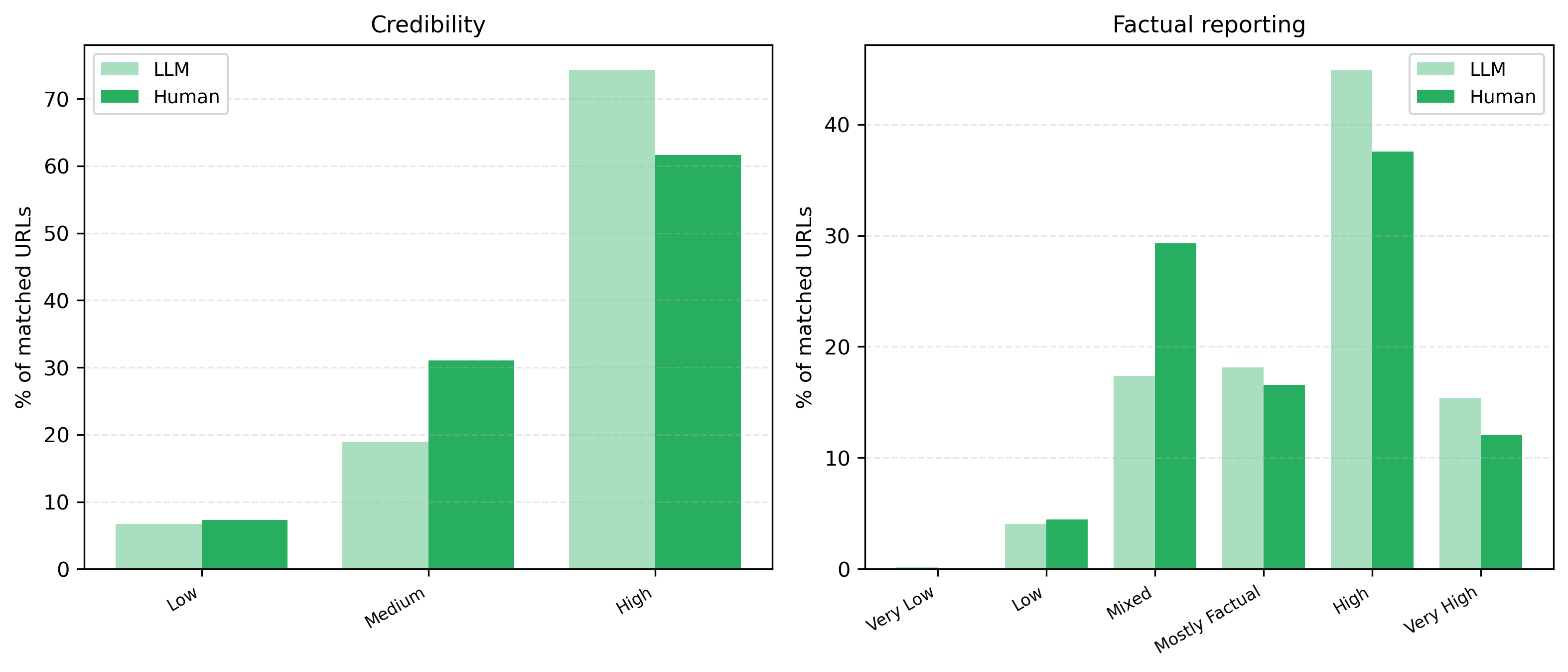}
  \caption{Credibility and factual reporting of sources cited in LLM- and human-written notes. Bars show the percentage of MBFC-matched URLs in each credibility category (left) and factual reporting category (right), separately for LLM and human notes. LLM-cited sources are more concentrated in the highest credibility and factual reporting categories.}
  \label{fig:mbfc_source}
\end{figure}

Treating each matched citation as one observation, sources cited by LLM notes had higher average factual reporting than those cited by human notes (3.478 versus 3.207 on a 0--5 scale; Cohen's $d = 0.254$; $p<0.001$) and higher average credibility (1.660 versus 1.528 on a 0--2 scale; Cohen's $d = 0.224$; $p<0.001$). As shown in Figure~\ref{fig:mbfc_source}, LLM-cited sources were more concentrated in the highest credibility and factual reporting categories, whereas human-cited sources were more frequently classified in medium or mixed categories.

As a robustness check, we repeated this source quality analysis using the aggregate domain-quality ratings developed by \cite{lin2023domainquality}, which combines six expert rating sets, including MBFC, and ranges from 0 (lowest quality) to 1 (highest quality). Using the same URL extraction and domain-matching procedure, the aggregate measure matched 1,467 LLM-cited URLs (60.2\%) and 687 human-cited URLs (42.0\%). LLM-cited sources again had higher average quality scores (0.787 vs. 0.752, $t=4.21$, $p<0.001$).

\begin{table}[htbp!]
\caption{Top 10 domains cited in LLM notes, with comparison to human notes.}
\centering
\footnotesize
\begin{tabular}{clrr}
\toprule
Rank & Domain & \% LLM & \% human \\
\midrule
1 & reuters.com & 7.7 & 1.6 \\
2 & en.wikipedia.org & 7.2 & 7.6 \\
3 & instagram.com & 5.8 & 2.0 \\
4 & youtube.com & 5.3 & 3.9 \\
5 & x.com & 4.2 & 18.7 \\
6 & bbc.com & 4.2 & 1.1 \\
7 & snopes.com & 3.9 & 0.7 \\
8 & cnn.com & 2.4 & 0.8 \\
9 & facebook.com & 2.0 & 0.7 \\
10 & yahoo.com & 2.0 & 0.5 \\
\bottomrule
\end{tabular}
\label{tab:top10_llm_domains}
\end{table}

\begin{table}[htbp]
\caption{Top 10 domains cited in human notes, with comparison to LLM notes.}
\centering
\footnotesize
\begin{tabular}{clrr}
\toprule
Rank & Domain & \% LLM & \% human \\
\midrule
1 & x.com & 4.2 & 18.7 \\
2 & en.wikipedia.org & 7.2 & 7.6 \\
3 & youtube.com & 5.3 & 3.9 \\
4 & x.com/grok & 0.4 & 2.6 \\
5 & instagram.com & 5.8 & 2.0 \\
6 & reuters.com & 7.7 & 1.6 \\
7 & theguardian.com & 1.2 & 1.6 \\
8 & t.co & 0.0 & 1.1 \\
9 & bbc.com & 4.2 & 1.1 \\
10 & share.google & 0.0 & 1.0 \\
\bottomrule
\end{tabular}
\label{tab:top10_human_domains}
\end{table}

\subsection{Rater-Provided Tags Suggest Better Sourcing and More Neutral Language}

To better understand the differences in helpfulness between LLM and human notes, we analyze the tags selected by raters when evaluating notes. Raters may optionally select helpful tags when assigning a rating of ``helpful'' or ``somewhat helpful'', and unhelpful tags when assigning a rating of ``unhelpful'' or ``somewhat helpful''. 88.1\% of the ratings have at least one tag. To prevent notes with more ratings from disproportionately influencing the estimates, we first calculate the selection rate for each tag within each note. We then average these note-level proportions within each tweet and writer type and estimate the LLM--human difference within tweets.

The tag patterns suggest that raters perceive LLM notes as better sourced and less biased or argumentative. LLM notes were more likely to be tagged as citing good sources: the tweet-weighted average was 58.2\% for LLM notes, compared with 51.2\% for human notes ($t=6.69$, $p<0.001$). LLM notes were also more likely to receive the ``helpfulUnbiasedLanguage'' tag (48.3\% versus 45.5\%; $t=2.90$, $p=0.004$). The unhelpful tags provide complementary evidence. LLM notes were less likely than human notes to be marked as argumentative or biased (15.4\% versus 22.3\%; $t=-5.36$, $p<0.001$), as containing opinion or speculation (22.4\% versus 26.7\%; $t=-2.90$, $p=0.004$), or as having missing or unreliable sources (21.8\% versus 26.5\%; $t=-3.31$, $p<0.001$). However, LLM notes were more likely to be tagged as not needed (36.6\% versus 25.7\%; $t=7.40$, $p<0.001$), a pattern that often occurred when the LLM writer interpreted a post too literally and treated jokes, satire, or expressions of personal opinion as misleading claims requiring correction. Overall, these exploratory results suggest that the rating advantage of LLM notes is associated primarily with greater perceived source quality and more neutral language, together with fewer concerns about bias, speculation, and unreliable sourcing. At the same time, the LLM writer exhibited a complementary limitation where it sometimes identified factual issues but failed to recognize the contextual and social cues that determine whether a correction is needed at all.

\begin{table}[htbp]
\centering
\caption{\textbf{Rater-provided tags by writer type.} Percentages report the share of eligible ratings selecting each tag. Rating-weighted columns pool all eligible ratings. Matched-tweet columns average note-level proportions within tweet and writer type and test the paired LLM--human difference within tweets.}
\label{tab:rater_tags_mechanism}
\scriptsize
\setlength{\tabcolsep}{4pt}
\begin{adjustbox}{max width=\linewidth}
\begin{tabular}{lrrrrrr}
\toprule
& \multicolumn{2}{c}{Rating-weighted (\%)} & \multicolumn{2}{c}{Matched-tweet (\%)} & \multicolumn{2}{c}{Matched-tweet difference} \\
\cmidrule(lr){2-3}\cmidrule(lr){4-5}\cmidrule(lr){6-7}
Tag & LLM & Human & LLM & Human &  $t$ & $p$ \\
\midrule
\multicolumn{7}{l}{\textit{Panel A. Helpful or somewhat-helpful ratings (AI $n=36{,}799$, human $n=48{,}229$; paired tweets $=740$)}} \\
Addresses claim & 70.4 & 70.5 & 69.0 & 68.3 & 0.73 & 0.46 \\
Important context & 67.7 & 67.3 & 66.3 & 65.4 & 0.95 & 0.34 \\
Clear & 65.5 & 66.6 & 64.9 & 64.9 & 0.04 & 0.97 \\
Good sources & 59.7 & 54.6 & 58.2 & 51.2 & 6.69 & $<0.001$\\
Unbiased language & 49.8 & 47.9 & 48.3 & 45.5 & 2.90 & 0.004 \\
Other & 8.2 & 7.7 & 8.1 & 8.8 & -1.27 & 0.20 \\
\midrule
\multicolumn{7}{l}{\textit{Panel B. Not-helpful or somewhat-helpful ratings (AI $n=9{,}723$, human $n=16{,}215$; paired tweets $=497$)}} \\
Missing key points & 44.5 & 44.2 & 38.4 & 39.0 & -0.44 & 0.66 \\
Note not needed & 42.5 & 26.0 & 36.6 & 25.7 & 7.40 & $<0.001$\\
Opinion/speculation & 22.1 & 31.8 & 22.4 & 26.7 & -2.90 & 0.004 \\
Missing/unreliable sources & 20.7 & 24.9 & 21.8 & 26.5 & -3.31 & $<0.001$\\
Incorrect & 22.7 & 22.9 & 24.0 & 22.6 & 0.99 & 0.32 \\
Argumentative/biased & 16.3 & 27.9 & 15.4 & 22.3 & -5.36 & $<0.001$\\
Irrelevant sources & 14.0 & 15.6 & 15.9 & 15.9 & 0.03 & 0.98 \\
Other & 10.9 & 8.3 & 14.4 & 9.2 & 4.46 & $<0.001$\\
Hard to understand & 6.0 & 8.3 & 8.8 & 9.5 & -0.64 & 0.52 \\
Spam/harassment/abuse & 4.8 & 7.4 & 5.0 & 5.5 & -0.74 & 0.46 \\
\bottomrule
\end{tabular}
\end{adjustbox}
\end{table}

\section{Discussion}

Our field deployment provides proof-of-concept evidence that LLM-generated Community Notes can achieve broad cross-ideological acceptance in a live platform environment. LLM-written notes received more positive ratings across the political spectrum, consistent with Community Notes' emphasis on cross-perspective consensus rather than simple accuracy or majority agreement. 
The topic-level heterogeneity we observe offers insight into where LLM fact-checking is most and least effective. The LLM writer's strongest advantages appear on health and medicine claims and on conspiracy theories and pseudoscience, domains where authoritative sources are abundant and factual grounding is relatively clear-cut. By contrast, the LLM shows minimal advantage on posts about AI-generated content, where detection often requires specialized capabilities that our LLM writer does not access. Tools such as reverse image search or Google's SynthID could be integrated into future pipelines to improve detection of AI-generated content. We also find that raters perceived LLM notes as better sourced and more neutrally written, with fewer concerns about bias, argumentativeness, speculation, and unreliable sourcing, while also identifying a recurring failure mode that the LLM sometimes proposed unnecessary notes. More broadly, our results reflect the performance of a single LLM writer pipeline; other AI writer implementations on the platform may differ in topic-specific strengths and overall performance.

These findings should be interpreted alongside important platform-specific confounds. LLM notes currently face a structural timing disadvantage: they can be written only after sufficient users flag a post, meaning they are systematically submitted later and accumulate fewer ratings. Because the Community Notes algorithm penalizes notes with fewer ratings, this timing disadvantage can deflate LLM notes' helpfulness scores. Our common-rater note-level analysis and robustness checks address this issue in several ways, suggesting that LLM notes receive higher helpfulness scores when differential exposure is controlled. Together, the pattern indicates that the core quality signal favors LLM notes, while timing-induced exposure effects obscure this signal in aggregated platform metrics.

Our findings have broader implications for the role of LLMs in public information ecosystems \citep{li2025}. To our knowledge, this study is the first and the largest field evaluation of AI-generated fact-checking on any social media platform, made possible by the transparent data access provided by Community Notes. This open data enables a truly ecological and valid assessment: rather than relying on recruited evaluators or proxy metrics, we examine how a diverse population of real platform users respond to AI-generated content under natural conditions.

Community Notes provides a distinctive testing ground for AI fact-checking: notes are evaluated by a diverse, self-selected rater population on a live platform. The fact that LLMs can meet this standard at scale suggests that AI-augmented fact-checking is viable not merely in controlled offline settings, but under the conditions of real social media platforms. This is particularly consequential given the scale mismatch between misinformation production and human fact-checking capacity. Human contributors, however motivated, are limited in the volume of content they can monitor and annotate. LLM writers face no such constraint, and our results suggest that scaling up AI contributions need not come at the cost of quality or cross-ideological acceptability. At the same time, LLM and human contributors have different strengths that are not easily substituted \citep{li2025}. LLMs are good at rapidly synthesizing widely available information and generating contextual notes at scale, while human contributors are better equipped to handle novel situations, niche topics, rapidly evolving events, and cases that require deep social or cultural context. By having LLM- and human-written notes compete within the same rating system, the most helpful notes, regardless of origin, rise to the top. Platforms and researchers designing AI-augmented fact-checking systems should consider how to harness this complementarity that enhances overall note quality rather than treating AI deployment as a replacement for human community participation.

Beyond the respective strengths of human and AI contributors, the rapid expansion of AI participation also raises broader questions about the evolution of the fact-checking ecosystem. Recent descriptive evidence suggests that AI community notes now constitute a substantial share of notes on the platform and are produced by a small number of operators \citep{mantzarlis2026aibots}. Our study is scoped to a single writer whose pipeline we can document, so we do not characterize or compare with writers whose operations are largely unobservable (see \cite{gong2026characterizingaifactcheckerscontributions} for a descriptive analysis of all AI writers on the platform). But these developments underscore the need to evaluate AI fact-checking as a broader ecosystem, such as how it affects the diversity and coverage of fact-checking and the continued participation of human writers and raters. 

\newpage
\section{Data and Methods}

\subsection{Dataset}

Our analysis dataset comprised 2,946 Community Notes written between November 1, 2025 and January 31, 2026, targeting 1,597 unique tweets, with 108,169 ratings from 42,521 unique raters. The sample included 1,614 notes written by our LLM writer and 1,332 notes written by human users on posts in the same deployment sample.\footnote{The sample size was determined by the available budget for LLM API usage. 17 posts have more than one note from our LLM writer because we operated multiple API accounts, and a minor bug once allowed different accounts to submit notes on the same post.} Notes from other AI writers were excluded. We applied two quality filters: excluding media notes and retaining only ratings from raters with a valid rater factor.

All notes and ratings come from the Community Notes public data releases, and note helpfulness scores and rater factors were computed using the Community Notes open-sourced scoring algorithms \citep{communitynotes_ranking}. All human-subject data were de-identified prior to use in this study. Contributor identities are pseudonymized by the platform. We analyzed only these pseudonymized identifiers, report results in aggregate, and did not attempt to re-identify any contributor.

\subsection{Analysis and Testing Strategy}

Our analyses operate at two levels. The rating-level analysis estimates a linear mixed-effects model predicting individual rating scores; p-values on regression coefficients are interpreted without multiple-testing correction as they are jointly estimated within a single model. Subgroup analyses by tweet modality and topic re-estimate this specification within subsets and are treated as exploratory. At the note level, our primary analysis uses the common-rater design, in which we test three outcomes---note helpfulness score, CRH status, and CRNH status---using a linear mixed-effects model with tweet random intercepts to account for within-tweet dependence among notes targeting the same post. We apply Benjamini--Hochberg correction across these three tests to control the false discovery rate.

\subsection{Note Writing Pipeline}

Upon retrieving a potentially misleading post via the AI writer API, we compile its context by concatenating the post's creation timestamp, primary text, and any text and image descriptions from quoted or replied-to posts, if applicable. To enable multimodal note writing, the LLM writer is directly provided with images or video thumbnails associated with the target post. Next, we use Grok-4-fast with web search and X search capabilities to conduct information research and gather relevant evidence for the note-writing process. Grok was chosen for its ability to surface X posts and understand video content from X, in addition to performing web search. After evidence collection, the LLM writer (GPT-5-mini) decides whether a note should be written for the post. A note may not be written if the post is unlikely to be perceived as misleading (29.0\% of all retrieved posts) or if there is insufficient evidence (6.9\%). This triage step acts as a safeguard against generating spam notes on non-misleading posts. Once the decision to write a note is made, the LLM (GPT-5-mini) writes the Community Note using the post content and Grok research outputs, following Community Notes guidelines. The note then passes through URL validity, length, and quality checks using the Community Notes ClaimOpinion model\footnote{\href{https://docs.x.com/x-api/community-notes/evaluate-a-community-note}{https://docs.x.com/x-api/community-notes/evaluate-a-community-note}} prior to submission. The pipeline did not query, retrieve, or pass existing Community Notes, ratings, or statuses as inputs, so notes were written independently of any other notes that may already have been present. 

We refined this pipeline iteratively during the development based on inspection of generated notes and their recurring failure modes: separating evidence retrieval from note generation makes it possible to audit how evidence is collected, selected, and incorporated; the retrieval prompt explicitly requests sources spanning the ideological spectrum to promote cross-perspective acceptability; the retrieval stage incorporates contextual information about the post author and audience reactions to reduce the risk of misclassifying satire or opinion as factually false based on a literal interpretation; the triage stage accounts for the fact that not all retrieved posts are actually misleading or warrant a note; post-generation checks for URL validity, note length, and the platform's ClaimOpinion score help filter residual issues, such as opinionated phrasing, that prompting alone cannot fully eliminate.

\section*{Acknowledgements}
We thank Dean Eckles, Martin Saveski, and Soham De for insightful conversations and feedback on this work. We thank Brad Miller, Jay Baxter, and Keith Coleman for valuable discussions on Community Notes and AI fact-checking.

We used AI to assist with writing and debugging analysis code and to provide feedback on drafts of this paper. All analyses, interpretations, and final writing decisions were made by the authors.
\clearpage
\bibliographystyle{plainnat}
\bibliography{references}

\clearpage
\appendix
\setcounter{secnumdepth}{2}

{\noindent\LARGE\textbf{Supplementary Information}}
\vspace{1em}

\renewcommand{\thefigure}{S\arabic{figure}}
\renewcommand{\thetable}{S\arabic{table}}
\setcounter{figure}{0}
\setcounter{table}{0}

\section{Supplementary Rating-Level Analyses}
\subsection{Within-Rater Pairwise Comparison of Ratings}
\label{app:within_rater}

As a complement to the rating-level analysis, we conducted a pairwise comparison restricted to raters who evaluated both an LLM and a human note on the same tweet. We constructed 1,186 unique LLM--human note pairs, yielding 21,978 rater--pair observations from 10,345 raters. For each observation, we encoded a win/loss/tie outcome based on whether the rater's rating for the LLM note was better than their rating for the human note. Across all rater–pair observations, 71.0\% were ties, 15.8\% were LLM wins, and 13.2\% were human wins. Excluding ties, a Bradley-Terry logistic model with standard errors clustered by rater indicates that the LLM note was preferred in 54.4\% of non-tied comparisons ($\beta = 0.178$, SE $= 0.035$, $z = 5.10$, $p < 0.001$; OR $= 1.19$), significantly above the 50\% chance level.

\subsection{Temporal Dynamics of Ratings}
\label{app:rating_temporal}

Ratings may depend on when they are submitted relative to a note's creation. We introduce a temporal structure into the rating-level analysis by extending primary specification in Equation~\ref{eq:main} to include the log-transformed elapsed time between note creation and rating submission, along with its interaction with LLM authorship. We measure elapsed time in hours and transform it as $\log(1+\text{elapsed hours})$ to account for its skewed distribution. We then mean-center the
log-transformed measure because this makes the coefficient on AI interpretable as the estimated
LLM--human difference at the average logged elapsed time in the sample.

After introducing this temporal structure, the estimated LLM advantage remains nearly identical (coefficient = $0.105$, $p<0.001$, see table \ref{tab:temporal_rating}). Ratings of human-written notes become less positive as logged elapsed time increases (coefficient $=-0.013$, $p<0.001$). The interaction between LLM authorship and logged elapsed time is positive (coefficient $=0.006$, $p=0.003$), indicating that ratings of LLM-written notes decline less steeply over time. The coefficients describing heterogeneity by rater ideology also remain near identical to those in the primary specification. We note that this analysis is descriptive and therefore show that the primary findings are robust to observed elapsed time, rather than causal, as rating timing is endogenous. 

\begin{table}[htbp]
\centering
\caption{\textbf{Rating-level model incorporating elapsed time.}
Elapsed time is measured as the mean-centered value of
$\log(1+\text{hours between note creation and rating submission})$. The model includes note and rater
random intercepts.}
\label{tab:temporal_rating}
\small
\begin{adjustbox}{max width=\linewidth}
\begin{tabular}{lrrr}
\toprule
 & Coefficient & SE & $p$ \\
\midrule
AI & 0.105 & 0.009 & $<0.001$ \\
Rater ideology & $-0.016$ & 0.004 & $<0.001$ \\
Rater ideology$^{2}$ & $-0.174$ & 0.010 & $<0.001$ \\
Centered log time since note creation
& $-0.013$ & 0.001 & $<0.001$ \\
AI $\times$ rater ideology
& $-0.087$ & 0.005 & $<0.001$ \\
AI $\times$ rater ideology$^{2}$
& $-0.191$ & 0.013 & $<0.001$ \\
AI $\times$ centered log time
& 0.006 & 0.002 & 0.003 \\
\midrule
SD: note intercept & 0.196 & & \\
SD: rater intercept & 0.159 & & \\
Residual SD & 0.304 & & \\
Observations & 108,169 & & \\
\bottomrule
\end{tabular}
\end{adjustbox}
\end{table}

\subsection{Textual Features as Predictors of Ratings}
\label{app:word_url_count}

As presented in the paper, LLM and human notes differ in several observable textual characteristics, including length, citation count, readability, valence, and toxicity. These differences could partly account for the estimated association between LLM authorship and rating helpfulness. We therefore extend the primary rating-level specification in Equation~\ref{eq:main} to include standardized measures of these features. URLs are removed before calculating the language-based measures so that citation formatting does not mechanically affect them. All features are standardized using their pooled means and standard deviations across unique notes.

We estimate two additional specifications. The first adds standardized textual features to the primary model, and the second interacts all features with LLM authorship, allowing these associations to differ between LLM- and human-written notes. The estimated LLM advantage remains similar across these specifications and the interactions between LLM authorship and rater ideology remain essentially unchanged (see table \ref{tab:rating_text_features}). Thus, the measured textual differences account for only a modest portion of the estimated LLM rating advantage. We find only citation count and toxicity significantly predict ratings. With the exception of citation count, none of the interactions with LLM authorship is statistically significant. Citation count is positively associated with ratings of human-written notes ($\beta=0.050$, $p<0.001$), but its negative interaction with LLM authorship ($\beta=-0.052$, $p<0.001$) implies an
approximately zero association among LLM notes. 

\begin{table}[htbp]
\centering
\caption{\textbf{Rating-level models adjusting for observable textual features.} The outcome is the individual rating score. All textual features are standardized using their pooled means and standard deviations across unique LLM- and human-written notes. 18 notes are dropped because they are too short for the Flesch-Kincaid scoring.}
\label{tab:rating_text_features}
\small
\begin{adjustbox}{max width=\linewidth}
\begin{tabular}{lcc}
\toprule
& \multicolumn{2}{c}{Rating score} \\
\cmidrule(lr){2-3}
& Additive controls
& Interactions with LLM \\
\midrule

Intercept
& 0.795***
& 0.795*** \\
& (0.007)
& (0.007) \\

LLM-written note
& 0.088***
& 0.099*** \\
& (0.010)
& (0.010) \\

Rater ideology
& $-0.016$***
& $-0.016$*** \\
& (0.004)
& (0.004) \\

Rater ideology$^{2}$
& $-0.175$***
& $-0.175$*** \\
& (0.010)
& (0.010) \\

Standardized word count
& $-0.006$
& $-0.002$ \\
& (0.005)
& (0.005) \\

Standardized URL count
& 0.039***
& 0.050*** \\
& (0.004)
& (0.005) \\

Standardized Flesch--Kincaid grade
& 0.001
& $-0.008$ \\
& (0.005)
& (0.006) \\

Standardized toxicity
& $-0.019$***
& $-0.022$*** \\
& (0.004)
& (0.004) \\

Standardized valence
& $-0.005$
& $-0.003$ \\
& (0.004)
& (0.006) \\

LLM $\times$ rater ideology
& $-0.086$***
& $-0.086$*** \\
& (0.005)
& (0.005) \\

LLM $\times$ rater ideology$^{2}$
& $-0.190$***
& $-0.190$*** \\
& (0.013)
& (0.013) \\

LLM $\times$ standardized word count
& ---
& $-0.017$ \\
&
& (0.013) \\

LLM $\times$ standardized URL count
& ---
& $-0.052$*** \\
&
& (0.010) \\

LLM $\times$ standardized Flesch--Kincaid grade
& ---
& 0.011 \\
&
& (0.009) \\

LLM $\times$ standardized toxicity
& ---
& 0.014 \\
&
& (0.010) \\

LLM $\times$ standardized valence
& ---
& 0.001 \\
&
& (0.009) \\

\midrule
SD: rater intercept
& 0.160
& 0.160 \\

SD: note intercept
& 0.190
& 0.188 \\

Residual SD
& 0.304
& 0.304 \\

Observations
& 107,637
& 107,637 \\

\bottomrule
\end{tabular}
\end{adjustbox}

\par\vspace{4pt}
{\footnotesize\noindent
Standard errors are shown in parentheses.
* $p<0.05$, ** $p<0.01$, *** $p<0.001$.}
\end{table}

\subsection{Associations Between Frequently Cited Domains and Ratings}
We conducted an exploratory rating-level analysis examining whether citations to frequently referenced domains were associated with ratings and whether these associations differed between LLM- and human-written notes. We jointly added indicators for the commonly cited domains to the primary rating-level specification and controlled for the total number of URLs cited by each note. We then interacted LLM authorship with domains cited by at least 15 notes of each writer type.

In the additive model, citations to \texttt{x.com} were positively associated with ratings, as were citations to Facebook and Reuters (X and Facebook posts were frequently referenced as the original sources of the claims or media found in misleading posts). The interaction model indicated that the positive association with \texttt{x.com} citations was concentrated among human-written notes: the coefficient among human notes was positive and significant, while the interaction with LLM authorship implied an approximately zero association among LLM-written notes. None of the other source-by-writer interactions was statistically significant. Because source selection is endogenous, these estimates are descriptive and should not be interpreted as causal effects of citing particular domains.

\begin{table}[htbp]
\centering
\caption{\textbf{Associations between cited domains and ratings.}
The outcome is the individual rating score. Domain indicators are entered
jointly, and both models control for the standardized total number of URLs
cited by the note.}
\label{tab:rating_cited_domains}
\small
\begin{adjustbox}{max width=\linewidth}
\begin{tabular}{lcc}
\toprule
& \multicolumn{2}{c}{Rating score} \\
\cmidrule(lr){2-3}
& Additive domain indicators
& Interactions with LLM \\
\midrule

LLM-written note
& 0.094*** (0.009)
& 0.110*** (0.011) \\

Standardized URL count
& 0.033*** (0.004)
& 0.033*** (0.004) \\

\addlinespace
\multicolumn{3}{l}{\textit{Cited-domain indicators}} \\

\texttt{bbc.com}
& 0.017 (0.025)
& 0.053 (0.055) \\

\texttt{cnn.com}
& $-0.020$ (0.031)
& --- \\

\texttt{en.wikipedia.org}
& 0.012 (0.016)
& 0.036 (0.022) \\

\texttt{facebook.com}
& 0.092** (0.034)
& --- \\

\texttt{instagram.com}
& 0.039 (0.021)
& 0.052 (0.042) \\

\texttt{reuters.com}
& 0.038* (0.019)
& 0.080 (0.045) \\

\texttt{share.google}
& 0.023 (0.059)
& --- \\

\texttt{snopes.com}
& 0.030 (0.027)
& --- \\

\texttt{t.co}
& 0.003 (0.056)
& --- \\

\texttt{theguardian.com}
& $-0.016$ (0.036)
& $-0.037$ (0.049) \\

\texttt{x.com}
& 0.064*** (0.014)
& 0.079*** (0.016) \\

\texttt{x.com/grok}
& 0.016 (0.035)
& --- \\

\texttt{yahoo.com}
& 0.045 (0.037)
& --- \\

\texttt{youtube.com}
& 0.032 (0.021)
& 0.061 (0.033) \\

\addlinespace
\multicolumn{3}{l}{\textit{Interactions with LLM authorship}} \\

LLM $\times$ \texttt{bbc.com}
& ---
& $-0.044$ (0.062) \\

LLM $\times$ \texttt{en.wikipedia.org}
& ---
& $-0.052$ (0.031) \\

LLM $\times$ \texttt{instagram.com}
& ---
& $-0.016$ (0.048) \\

LLM $\times$ \texttt{reuters.com}
& ---
& $-0.056$ (0.050) \\

LLM $\times$ \texttt{theguardian.com}
& ---
& 0.033 (0.072) \\

LLM $\times$ \texttt{x.com}
& ---
& $-0.074$* (0.034) \\

LLM $\times$ \texttt{youtube.com}
& ---
& $-0.052$ (0.042) \\

\midrule

Rater ideology controls
& Yes
& Yes \\

LLM $\times$ rater ideology controls
& Yes
& Yes \\

SD: rater intercept
& 0.160
& 0.160 \\

SD: note intercept
& 0.190
& 0.190 \\

Residual SD
& 0.304
& 0.304 \\

Observations
& 107,637
& 107,637 \\

\bottomrule
\end{tabular}
\end{adjustbox}

\par\vspace{4pt}
{\footnotesize\noindent
Standard errors are shown in parentheses. * $p<0.05$, ** $p<0.01$, *** $p<0.001$.}
\end{table}

\newpage
\section{Supplementary Note-Level Analyses}
\subsection{Observed Platform Outcomes in the Full Sample}
\label{app:full_sample}

We present note-level analyses of note helpfulness scores, CRH status, and CRNH status in the full sample. A higher proportion of human notes achieved CRH status than LLM notes (18.02\% vs.\ 13.07\%; AI coef.\ $= -0.058$, $z = -4.556$, $p < 0.001$). Among CRH notes, human notes reached that status faster (median: 5.90 hours vs.\ 7.38 hours for LLM notes; Mann--Whitney $U = 27{,}979$, $p = 0.054$). However, LLM notes had a lower rate of Currently Rated Not Helpful (CRNH) status (1.12\% vs.\ 4.13\% for human notes; AI coef.\ $= -0.027$, $z = -4.714$, $p < 0.001$) and similar average note helpfulness scores (mean: 0.25 vs.\ 0.24; AI coef.\ $= 0.007$, $z = 1.143$, $p = 0.253$). These results should be interpreted in light of substantial platform-specific confounds: the Community Notes algorithm used to compute these outcomes penalizes notes with few ratings, and LLM writers face a submission-timing disadvantage that causes them to accumulate fewer ratings.

To benchmark the LLM writer's performance against the distribution of individual human writers, we compare the CRH rate and hit rate. The CRH rate is the fraction of a writer's notes achieving CRH status, and the hit rate is defined as (CRH notes $-$ CRNH notes) / total notes to account for notes reaching unhelpful status. Our LLM writer achieves a CRH rate of 13.07\%, outranking 84.7\% of human writers on CRH rate (78.4\% among human writers who have written at least 30 notes), and a hit rate of 11.96\%, corresponding to the 85.4th percentile among human writers (79.0\% among human writers who have written at least 30 notes).

Tables~\ref{tab:A1_rating} and~\ref{tab:A2_notelevel} summarize the rating-level and note-level results for the full sample, respectively.

\begin{table}[htbp]
\centering
\caption{\textbf{Mean \% helpful and \% unhelpful ratings per note by rater ideology bucket (full sample).} $n$ denotes the number of notes receiving at least one rating from raters in that ideology bucket. 95\% CIs are computed across notes.}
\label{tab:A1_rating}
\small
\begin{adjustbox}{max width=\linewidth}
\begin{tabular}{@{}lcccc@{}}
\toprule
& \multicolumn{2}{c}{\% Helpful, mean {[}95\% CI{]}} & \multicolumn{2}{c}{\% Unhelpful, mean {[}95\% CI{]}} \\
\cmidrule(lr){2-3}\cmidrule(lr){4-5}
Rater Ideology & LLM & Human & LLM & Human \\
\midrule
Left ($n$: 1,300 / 1,183)
  & 79.55 {[}77.83, 81.27{]} & 68.10 {[}65.95, 70.24{]}
  & 17.98 {[}16.33, 19.63{]} & 28.72 {[}26.61, 30.83{]} \\[3pt]
Neutral ($n$: 1,174 / 1,049)
  & 89.58 {[}88.32, 90.84{]} & 76.83 {[}74.86, 78.79{]}
  & \phantom{0}7.60 {[}6.52, 8.69{]} & 18.40 {[}16.56, 20.25{]} \\[3pt]
Right ($n$: 1,300 / 1,196)
  & 73.13 {[}71.32, 74.93{]} & 67.46 {[}65.52, 69.41{]}
  & 23.65 {[}21.92, 25.38{]} & 28.58 {[}26.68, 30.48{]} \\
\bottomrule
\end{tabular}
\end{adjustbox}
\end{table}

\begin{table}[htbp]
\centering
\caption{\textbf{Note-level comparisons in the full sample.} CRH, CRNH, and helpfulness score are analyzed using linear mixed-effects models (LMMs), and the reported coefficient is the estimated effect of the AI-note indicator. Number of ratings and time to CRH are compared using two-sided Mann--Whitney $U$ tests.}
\label{tab:A2_notelevel}
\small
\begin{adjustbox}{max width=\linewidth}
\begin{tabular}{lccccccc}
\toprule
Metric & LLM & Human & Test & AI coef. & Statistic & $p$ \\
\midrule
$N$ (total notes) & 1,614 & 1,332 & --- & --- & --- & --- \\

CRH rate & 13.07\% & 18.02\% & LMM & $-0.058$ & $z=-4.56$ & $<0.001$ \\
CRNH rate & 1.12\% & 4.13\% & LMM & $-0.027$ & $z=-4.71$ & $<0.001$ \\

Helpfulness score & 0.25 & 0.24 & LMM & $0.007$ & $z=1.14$ & 0.253 \\
\quad ($N$ with scores: 1,243 / 1,130) & & & & & & \\

Num.\ ratings, median & 22 & 51 & Mann--Whitney $U$ & --- & $U=740{,}798$ & $<0.001$ \\

Time to CRH, hrs, median & 7.38 & 5.90 & Mann--Whitney $U$ & --- & $U=27{,}979$ & 0.054 \\
\quad ($N$ CRH notes: 211 / 240) & & & & & & \\

\bottomrule
\end{tabular}
\end{adjustbox}
\end{table}

\subsection{Sensitivity to Minimum Rating Thresholds}
\label{app:robustness}

In addition to the note-level analysis with common raters, we conducted two robustness checks to address potential confounds in note-level outcomes. To address the concern that notes with fewer ratings have overly penalized helpfulness scores, we examine note-level outcomes after requiring notes to have at least 10, 30, or 50 ratings. Table~\ref{tab:A3_notelevel_thresholds} reports the results. At the 10-rating threshold, LLM notes remain less likely to reach CRH status than human notes (18.7\% vs. 20.9\%; AI coefficient $=-0.038$, $p=0.015$). This difference attenuates as the threshold increases and is no longer statistically distinguishable at either 30 ratings (27.9\% versus 27.0\%; AI coefficient $=-0.006$, $p=0.751$) or 50 ratings (29.4\% versus 28.8\%; AI coefficient $=-0.005$, $p=0.824$). Across all three thresholds, LLM notes are significantly less likely to reach CRNH status. LLM notes also have higher average helpfulness scores at each threshold.

\subsection{Creation-Time--Matched Note Outcomes}
To account for differential note exposure due to submission timing, we conduct creation-time--matched analyses. Results are presented in table~\ref{tab:timing_matched_notelevel}. We retain human notes created within $\pm$30, $\pm$60, or $\pm$90 minutes of the LLM note created on the same post, under the assumption that notes written close in time have similar exposure. Because LLM and human notes are often written hours apart, matching rates are low (6.3\%, 11.8\%, and 15.9\%, respectively) and sample sizes are substantially reduced. Within the $\pm$60-minute window ($n = 405$ notes; LLM $= 190$, human $= 215$), the CRH gap is narrow and non-significant (LLM $= 21.58\%$ vs.\ human $= 22.79\%$; AI coef.\ $= -0.019$, $z = -0.610$, $p = 0.542$). The difference in average note helpfulness scores is larger but not statistically significant (LLM $= 0.28$ vs.\ human $= 0.25$; AI coef.\ $= 0.023$, $z = 1.703$, $p = 0.089$). In this matched sample, LLM notes reach CRH status faster (median $= 4.61$ hours vs.\ $5.55$ hours).

\begin{table}[htbp]
\centering
\caption{\textbf{Note-level comparisons across minimum-rating
thresholds.} Each sub-table restricts the analysis to notes receiving at
least the indicated number of ratings. CRH, CRNH, and helpfulness score are
analyzed using linear mixed-effects models (LMMs). Number of ratings and time to CRH are compared using two-sided
Mann--Whitney $U$ tests.}
\label{tab:A3_notelevel_thresholds}
\small

\begin{subtable}{\linewidth}
\centering
\begin{adjustbox}{max width=\linewidth}
\begin{tabular}{lcccccc}
\toprule
Metric & LLM & Human & Test & AI coef. & Statistic & $p$ \\
\midrule
$N$ (total notes)
& 1,127 & 1,146 & --- & --- & --- & --- \\

CRH rate
& 18.72\% & 20.94\%
& LMM & $-0.038$ & $z=-2.44$ & 0.015 \\

CRNH rate
& 1.51\% & 4.28\%
& LMM & $-0.027$ & $z=-3.79$ & $<0.001$ \\

Helpfulness score
& 0.27 & 0.25
& LMM & $0.014$ & $z=2.14$ & 0.034 \\

Num.\ ratings, median
& 39.00
& 64.50
& Mann--Whitney $U$ & ---
& $U=493{,}006.50$ & $<0.001$ \\

Time to CRH, hrs, median
& 7.38
& 5.90
& Mann--Whitney $U$ & ---
& $U=27{,}979.00$ & 0.054 \\
\quad ($N$ CRH notes: 211 / 240)
& & & & & & \\

\bottomrule
\end{tabular}
\end{adjustbox}
\caption{At least 10 ratings. Retained 1,127/1,614 LLM notes
(69.8\%) and 1,146/1,332 human notes (86.0\%).}
\label{tab:threshold_10}
\end{subtable}

\vspace{1em}

\begin{subtable}{\linewidth}
\centering
\begin{adjustbox}{max width=\linewidth}
\begin{tabular}{lcccccc}
\toprule
Metric & LLM & Human & Test & AI coef. & Statistic & $p$ \\
\midrule
$N$ (total notes)
& 680 & 858 & --- & --- & --- & --- \\

CRH rate
& 27.94\% & 27.04\%
& LMM & $-0.006$ & $z=-0.32$ & 0.751 \\

CRNH rate
& 0.59\% & 4.55\%
& LMM & $-0.035$ & $z=-4.12$ & $<0.001$ \\

Helpfulness score
& 0.31 & 0.28
& LMM & $0.019$ & $z=2.40$ & 0.016 \\

Num.\ ratings, median
& 67.00
& 91.50
& Mann--Whitney $U$ & ---
& $U=236{,}530.50$ & $<0.001$ \\

Time to CRH, hrs, median
& 7.16
& 5.78
& Mann--Whitney $U$ & ---
& $U=24{,}102.00$ & 0.098 \\
\quad ($N$ CRH notes: 190 / 232)
& & & & & & \\

\bottomrule
\end{tabular}
\end{adjustbox}
\caption{At least 30 ratings. Retained 680/1,614 LLM notes
(42.1\%) and 858/1,332 human notes (64.4\%).}
\label{tab:threshold_30}
\end{subtable}

\vspace{1em}

\begin{subtable}{\linewidth}
\centering
\begin{adjustbox}{max width=\linewidth}
\begin{tabular}{lcccccc}
\toprule
Metric & LLM & Human & Test & AI coef. & Statistic & $p$ \\
\midrule
$N$ (total notes)
& 466 & 676 & --- & --- & --- & --- \\

CRH rate
& 29.40\% & 28.85\%
& LMM & $-0.005$ & $z=-0.22$ & 0.824 \\

CRNH rate
& 0.86\% & 4.44\%
& LMM & $-0.030$ & $z=-2.94$ & 0.003 \\

Helpfulness score
& 0.31 & 0.29
& LMM & $0.017$ & $z=1.86$ & 0.063 \\

Num.\ ratings, median
& 97.00
& 116.50
& Mann--Whitney $U$ & ---
& $U=132{,}070.00$ & $<0.001$ \\

Time to CRH, hrs, median
& 6.47
& 5.51
& Mann--Whitney $U$ & ---
& $U=15{,}055.00$ & 0.049 \\
\quad ($N$ CRH notes: 137 / 195)
& & & & & & \\

\bottomrule
\end{tabular}
\end{adjustbox}
\caption{At least 50 ratings. Retained 466/1,614 LLM notes
(28.9\%) and 676/1,332 human notes (50.8\%).}
\label{tab:threshold_50}
\end{subtable}

\end{table}

\begin{table}[htbp]
\centering
\caption{\textbf{Note-level comparisons in creation-time--matched subsets.} Each sub-table restricts analysis to LLM notes matched to human notes within the indicated submission-time window. CRH, CRNH, and helpfulness score are analyzed using linear mixed-effects models (LMMs); number of ratings and time to CRH are compared using two-sided Mann--Whitney $U$ tests.}
\label{tab:timing_matched_notelevel}
\small

\begin{subtable}{\linewidth}
\centering
\begin{adjustbox}{max width=\linewidth}
\begin{tabular}{lcccccc}
\toprule
Metric & LLM & Human & Test & AI coef. & Statistic & $p$ \\
\midrule
$N$ (total notes) & 102 & 116 & --- & --- & --- & --- \\

CRH rate & 21.57\% & 20.69\% & LMM & $0.004$ & $z=0.09$ & 0.929 \\
CRNH rate & 1.96\% & 6.90\% & LMM & $-0.049$ & $z=-1.74$ & 0.082 \\

Helpfulness score & 0.28 & 0.24 & LMM & $0.041$ & $z=2.13$ & 0.033 \\
\quad ($N$ with scores: 96 / 109) & & & & & & \\

Num.\ ratings, median & 64.00 & 55.00 & Mann--Whitney $U$ & --- & $U=6{,}491.5$ & 0.216 \\

Time to CRH, hrs, median & 6.01 & 5.67 & Mann--Whitney $U$ & --- & $U=288$ & 0.605 \\
\quad ($N$ CRH notes: 22 / 24) & & & & & & \\

\bottomrule
\end{tabular}
\end{adjustbox}
\caption{\(\pm 30\)-minute window. Matched 102/1,614 LLM notes (6.3\%).}
\label{tab:timing_30min}
\end{subtable}

\vspace{1em}

\begin{subtable}{\linewidth}
\centering
\begin{adjustbox}{max width=\linewidth}
\begin{tabular}{lcccccc}
\toprule
Metric & LLM & Human & Test & AI coef. & Statistic & $p$ \\
\midrule
$N$ (total notes) & 190 & 215 & --- & --- & --- & --- \\

CRH rate & 21.58\% & 22.79\% & LMM & $-0.019$ & $z=-0.61$ & 0.542 \\
CRNH rate & 2.11\% & 6.05\% & LMM & $-0.039$ & $z=-1.98$ & 0.048 \\

Helpfulness score & 0.28 & 0.25 & LMM & $0.023$ & $z=1.70$ & 0.089 \\
\quad ($N$ with scores: 178 / 193) & & & & & & \\

Num.\ ratings, median & 54.00 & 57.00 & Mann--Whitney $U$ & --- & $U=21{,}064.5$ & 0.587 \\

Time to CRH, hrs, median & 4.61 & 5.55 & Mann--Whitney $U$ & --- & $U=950$ & 0.662 \\
\quad ($N$ CRH notes: 41 / 49) & & & & & & \\

\bottomrule
\end{tabular}
\end{adjustbox}
\caption{\(\pm 60\)-minute window. Matched 190/1,614 LLM notes (11.8\%).}
\label{tab:timing_60min}
\end{subtable}

\vspace{1em}

\begin{subtable}{\linewidth}
\centering
\begin{adjustbox}{max width=\linewidth}
\begin{tabular}{lcccccc}
\toprule
Metric & LLM & Human & Test & AI coef. & Statistic & $p$ \\
\midrule
$N$ (total notes) & 257 & 313 & --- & --- & --- & --- \\

CRH rate & 19.46\% & 20.13\% & LMM & $-0.013$ & $z=-0.47$ & 0.638 \\
CRNH rate & 1.56\% & 5.43\% & LMM & $-0.037$ & $z=-2.37$ & 0.018 \\

Helpfulness score & 0.27 & 0.24 & LMM & $0.027$ & $z=2.37$ & 0.018 \\
\quad ($N$ with scores: 240 / 285) & & & & & & \\

Num.\ ratings, median & 57.00 & 59.00 & Mann--Whitney $U$ & --- & $U=40{,}772$ & 0.778 \\

Time to CRH, hrs, median & 4.96 & 5.46 & Mann--Whitney $U$ & --- & $U=1{,}562$ & 0.942 \\
\quad ($N$ CRH notes: 50 / 63) & & & & & & \\

\bottomrule
\end{tabular}
\end{adjustbox}
\caption{\(\pm 90\)-minute window. Matched 257/1,614 LLM notes (15.9\%).}
\label{tab:timing_90min}
\end{subtable}

\end{table}

\subsection{Within-Post Pairwise Comparison of Note Helpfulness Scores}
\label{app:within_post_scores}

As a note-level analog of the within-rater pairwise comparison, we compare the platform helpfulness scores (\texttt{coreNoteIntercept}) of LLM- and human-written notes targeting the same post. For each LLM--human note pair on the same post, we define an LLM win when the LLM note has the higher score and a human win when the human note has the higher score. Comparisons for which either note was not assigned an intercept due to insufficient number of ratings are excluded. Across all posts receiving both note types, 1,020 scored LLM--human pairs from 606 posts were available for comparison. The LLM writer has a win rate of 52.3\% (95\% CI $[49.2,55.3]$). The mean within-pair score difference was positive but small (LLM minus human $=0.029$).

Because posts vary in the number of LLM--human note pairs they contain, we also compute a post-weighted win rate. Each pair is weighted by the inverse of the number of pairs on its post, so every post contributes a total weight of one. The post-weighted LLM win rate was 46.4\% (95\% CI $[42.8,50.1]$), and the weighted mean score difference was approximately zero ($-0.001$).

\newpage
\subsection{Representativeness of Common Raters and Included Posts}
\label{app:rater_dist}
We use a common-rater design for our primary note-level analysis, in which outcomes are calculated using ratings from the subset of raters who evaluated all notes on a given tweet. To assess the representativeness of this analysis, we compare both the characteristics of these ``common raters'' and the topics of the included posts with those in the full sample.

\begin{figure}[H]
  \centering
  \includegraphics[width=\textwidth]{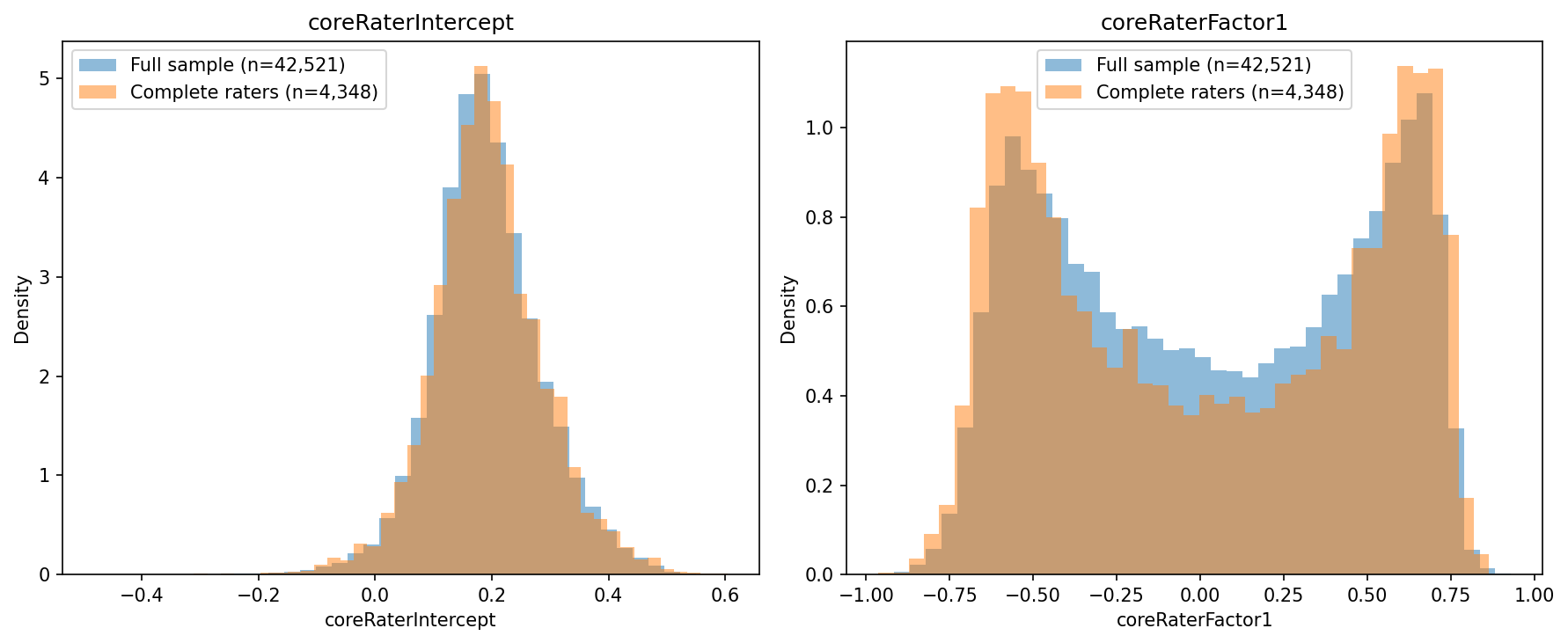}
  \caption{Distribution of rater characteristics for the full rater population vs.\ common raters who evaluated all notes on a given tweet. Left: \text{coreRaterIntercept} captures baseline helpfulness leniency. Right: \text{coreRaterFactor1} captures political leaning (negative = left-leaning, positive = right-leaning). The close overlap indicates that common raters are not systematically different from the overall rater population.}
  \label{fig:A1_rater_dist}
\end{figure}

Figure~\ref{fig:A1_rater_dist} compares the distribution of rater characteristics between the full rater population and the common-rater subset. The left panel shows the distribution of \text{coreRaterIntercept}, which measures a rater's baseline tendency to rate notes as helpful. The right panel shows the distribution of \text{coreRaterFactor1}, which captures political leaning inferred from historical rating patterns (negative = left-leaning, positive = right-leaning). Both distributions are highly similar across the two groups, suggesting that common raters do not systematically differ from the overall population in helpfulness leniency or political leaning.

We also compare the topic distribution of the 663 posts included in the common-rater analysis with that of all 1,597 posts in the full sample. The distributions do not differ significantly across the 11 topic categories ($\chi^2(10)=9.96$, $p=0.444$). Together, these comparisons indicate that restricting the analysis to common raters does not substantially alter the distributions of either key rater characteristics or post topics.

\subsection{Analyses Restricted to Posts with Both LLM and Human Notes}
\label{app:both_notes}
 
Our primary analyses use all 1,597 posts on which the LLM writer published a note, including 814 posts on which at least one human contributor also wrote a note. A natural concern is compositional: if posts annotated only by the LLM are systematically different from posts that human writers also chose to annotate, pooled comparisons could overstate the quality of LLM notes even with note and rater random effects. To address this concern directly, we replicate our analyses on the subset of 814 posts that received at least one LLM note and at least one human note, retaining all notes (821 LLM, 1,331 human) and all 94,347 ratings on these posts. The subset is topically representative of the full sample: a $\chi^2$ goodness-of-fit test against the full-sample topic distribution gives $\chi^2 = 9.3$, $df = 10$, $p = 0.51$.
 
Tables~\ref{tab:both_notes_rating}--\ref{tab:both_notes_hte} summarize the replication. All conclusions hold. Several analyses are unaffected by this restriction by construction and are therefore not repeated: the common-rater note-level analysis conditions on posts with at least one human note, the within-rater pairwise comparison requires both note types on the same post, and the creation-time--matched analyses match LLM notes to human notes on the same post.
 
\begin{table}[htbp]
\centering
\caption{\textbf{Rating-level results, full sample vs.\ both-note-types subset.} Top panel reports the AI coefficient (the LLM advantage at the center of the ideology scale) from the specification in Eq.~\ref{eq:main} and its robustness variants. Bottom panel reports mean \% helpful and \% unhelpful ratings per note by rater ideology group.}
\label{tab:both_notes_rating}
\small
\begin{adjustbox}{max width=\linewidth}
\begin{tabular}{lcc}
\toprule
& Full sample & Both-note posts \\
\midrule
\multicolumn{3}{l}{\textit{AI coefficient (SE)}} \\
Note + rater random intercepts (Eq.~\ref{eq:main}) & 0.104 (0.009)*** & 0.096 (0.011)*** \\
\midrule
\multicolumn{3}{l}{\textit{Mean \% helpful per note, LLM vs.\ human}} \\
Left-leaning raters & 79.6 vs.\ 68.1 & 77.3 vs.\ 68.1 \\
Neutral raters & 89.6 vs.\ 76.8 & 88.2 vs.\ 76.8 \\
Right-leaning raters & 73.1 vs.\ 67.5 & 72.3 vs.\ 67.5 \\
\midrule
\multicolumn{3}{l}{\textit{Mean \% unhelpful per note, LLM vs.\ human}} \\
Left-leaning raters & 18.0 vs.\ 28.7 & 20.5 vs.\ 28.7 \\
Neutral raters & \phantom{0}7.6 vs.\ 18.4 & \phantom{0}8.9 vs.\ 18.4 \\
Right-leaning raters & 23.7 vs.\ 28.6 & 24.5 vs.\ 28.5 \\
\bottomrule
\end{tabular}
\end{adjustbox}
\par\vspace{4pt}
{\footnotesize\noindent *** $p < 0.001$. Ratings: 108,169 (full) / 94,347 (subset).}
\end{table}
 
\begin{table}[htbp]
\centering
\caption{\textbf{Note-level outcomes, full sample vs.\ both-note-types subset.} CRH, CRNH, and helpfulness score analyzed with linear mixed-effects models with tweet random intercepts (Eq.~\ref{eq:2}); the AI coefficient is reported with $z$ and $p$.}
\label{tab:both_notes_notelevel}
\small
\begin{adjustbox}{max width=\linewidth}
\begin{tabular}{lcc}
\toprule
Metric & Full sample & Both-note posts \\
\midrule
$N$ notes (LLM / human) & 1,614 / 1,332 & 821 / 1,331 \\
CRH rate, LLM vs.\ human & 13.1\% vs.\ 18.0\% & 13.3\% vs.\ 18.0\% \\
\quad AI coef.\ & $-0.058$ ($z=-4.6$, $p<0.001$) & $-0.057$ ($z=-3.8$, $p<0.001$) \\
CRNH rate, LLM vs.\ human & 1.1\% vs.\ 4.1\% & 1.1\% vs.\ 4.1\% \\
\quad AI coef.\ & $-0.027$ ($z=-4.7$, $p<0.001$) & $-0.030$ ($z=-4.0$, $p<0.001$) \\
Helpfulness score, LLM vs.\ human & 0.25 vs.\ 0.24 & 0.26 vs.\ 0.24 \\
\quad AI coef.\ & $0.007$ ($p=0.25$) & $0.010$ ($p=0.17$) \\
Num.\ ratings, median & 22 vs.\ 51 & 29 vs.\ 51 \\
\midrule
\multicolumn{3}{l}{\textit{Subset with $\geq$30 ratings}} \\
$N$ notes (LLM / human) & 680 / 858 & 410 / 858 \\
CRH rate & 27.9\% vs.\ 27.0\% ($p=0.75$) & 23.9\% vs.\ 27.0\% ($p=0.40$) \\
CRNH rate & 0.6\% vs.\ 4.6\% ($p<0.001$) & 1.0\% vs.\ 4.6\% ($p=0.002$) \\
Helpfulness AI coef.\ & $0.019$ ($z=2.4$, $p=0.016$) & $0.016$ ($z=1.8$, $p=0.074$) \\
\bottomrule
\end{tabular}
\end{adjustbox}
\end{table}
 
\begin{table}[htbp]
\centering
\caption{\textbf{Subgroup AI coefficients by tweet modality and topic, full sample vs.\ both-note-types subset.} Estimates from the rating-level specification with note and rater random intercepts, fit within each subgroup; standard errors in parentheses.}
\label{tab:both_notes_hte}
\small
\begin{adjustbox}{max width=\linewidth}
\begin{tabular}{lcc}
\toprule
Subgroup & Full sample & Both-note posts \\
\midrule
Text-only & 0.127 (0.021) & 0.114 (0.025) \\
Image & 0.108 (0.015) & 0.104 (0.018) \\
Video & 0.086 (0.013) & 0.077 (0.016) \\
\midrule
Health \& Medicine & 0.170 (0.050) & 0.117 (0.071) \\
Conspiracy \& Pseudoscience & 0.145 (0.042) & 0.159 (0.055) \\
Social/Cultural Issues & 0.138 (0.023) & 0.137 (0.027) \\
Economy \& Finance & 0.135 (0.041) & 0.114 (0.051) \\
Politics \& Elections & 0.113 (0.019) & 0.095 (0.023) \\
Sports & 0.094 (0.038) & 0.109 (0.044) \\
Geopolitics \& Intl.\ Conflicts & 0.074 (0.022) & 0.074 (0.027) \\
Celebrity / Entertainment / Viral & 0.070 (0.020) & 0.059 (0.025) \\
Science \& Technology & 0.065 (0.041) & 0.040 (0.052) \\
AI generated content & 0.031 (0.038) & 0.016 (0.057) \\
Other / Miscellaneous & 0.005 (0.039) & 0.001 (0.046) \\
\bottomrule
\end{tabular}
\end{adjustbox}
\end{table}

We retain the full sample for our primary analyses for three reasons. First, the full-sample estimand matches our research question of evaluating an LLM writer in the field, which should measure how its notes are received across everything it actually wrote. Second, the head-to-head logic that motivates the restriction is already built into the analyses that carry our note-level conclusions, and we separately replicate the rating-level analysis in the both-note subset to address tweet-level selection directly. Third, the restriction discards 13\% of ratings and roughly half of LLM notes, reducing power without changing the substantive conclusions.

\section{Coverage, Selection, and Triage}
\subsection{Realized Post Coverage Across LLM and Human Writers}
\label{app:coverage_selection}

To characterize potential differences in the posts selected by our LLM writer and human writers, we compare posts that received only human notes, only notes from our LLM writer, or both note types during the same three-month study window. The deployment sample contains 783 LLM-only posts and 814 posts with both an LLM note and at least one human note. To construct a human-only comparison group, we identified 119,526 posts that received human notes during the study window but were not annotated by our LLM writer and sampled 3,800 posts from this pool. Of these, 3,058 remained live at the time of collection (July 28, 2026) and 1,628 were English-language posts, yielding the final human-only sample (we filtered to English posts for this comparison, because only English posts were available for the AI writers during the study window). 

We characterize each post by topic, modality, and the original poster's follower count. Topic and modality are the same LLM-assigned post-level labels used in the subgroup analyses. We group follower counts into three tiers: fewer than 100,000, 100,000--1 million, and more than 1 million followers. Follower counts were collected on July 28, 2026 rather than at the time the notes were written. 

\begin{figure}[H]
  \centering
  \includegraphics[width=\textwidth]{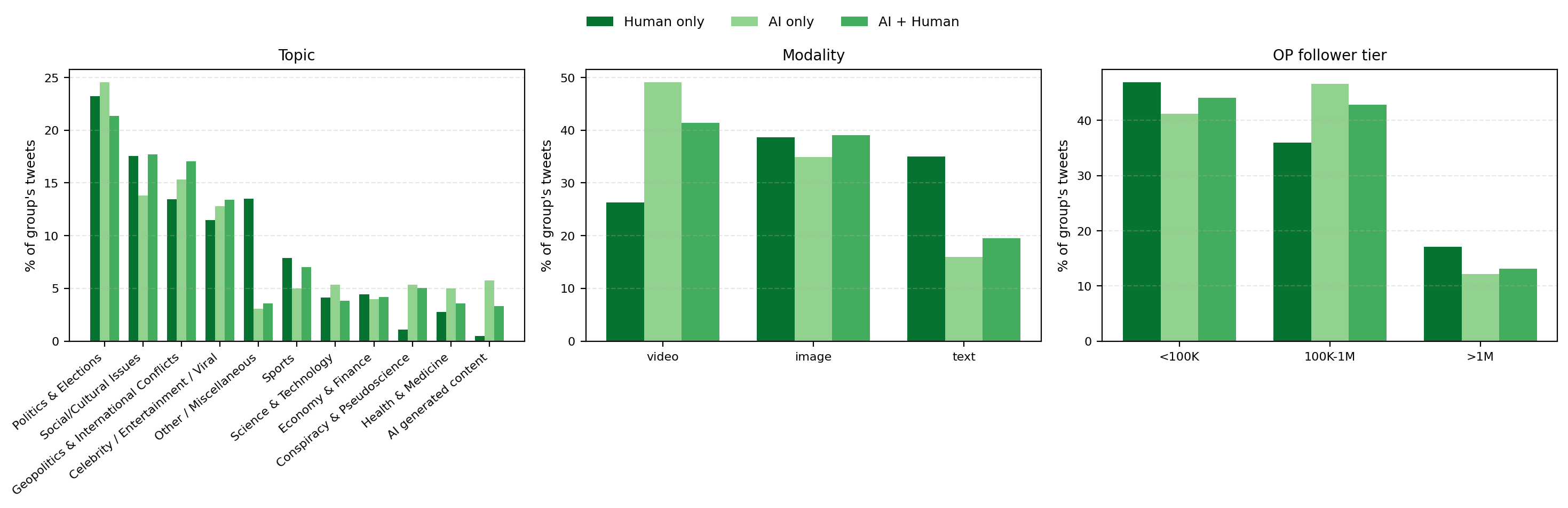}
  \caption{\textbf{Topic, modality, and OP follower tier by note-author group.} Human-only posts are English-language posts sampled from posts that received at least one human note but no note from our LLM writer during the three-month study window ($n=1{,}628$). LLM-only posts received a note from our writer but no human note ($n=783$), and LLM + human posts received both note types ($n=814$). Topic and modality are LLM-assigned post-level labels. Follower counts were from a later snapshot. Comparisons are descriptive.}
  \label{fig:coverage_selection}
\end{figure}

Figure~\ref{fig:coverage_selection} describes the coverage across the three groups. Politics and elections is the largest topic in all three groups, accounting for 21.4--24.6\% of posts. Relative to human-only posts, posts receiving an LLM note are more frequently about conspiracy and pseudoscience (5.0--5.4\% vs.\ 1.1\%) and AI-generated content (3.3--5.8\% vs.\ 0.5\%). Video posts account for 49.1\% of LLM-only posts and 41.4\% of posts with both note types, compared with 26.4\% of human-only posts. Conversely, text-only posts account for 16.0\%, 19.5\%, and 35.0\% of these groups, respectively. Follower-tier distributions are more similar across groups, although posts receiving an LLM note are somewhat more concentrated among accounts with 100,000--1 million followers. 

\subsection{Triage and Abstention Outcomes}
\label{app:abstention_outcomes}

Our LLM writer abstained for two distinct reasons: it judged that a post did not need a note, or it could not find sufficient evidence to support one. To characterize these decisions, we examine whether the abstained-on posts subsequently received notes from human contributors or from other AI writers. Because our writer abstained on these posts, all references to AI notes in this subsection refer to notes produced by other AI writers.

Table~\ref{tab:abstention_followup} reports the aggregate outcomes. Among posts classified as not needing a note, 38.55\% subsequently received at least one human-written note and 7.76\% received a human-written note that reached CRH status. Other AI writers annotated 9.98\% of these posts, and 1.99\% received an AI-written note that reached CRH status. Among posts for which our writer found insufficient evidence, 45.09\% subsequently received a human-written note and 12.22\% received a human-written CRH note. Other AI writers annotated 11.22\% of these posts, and 2.20\% received an AI-written CRH note.

\begin{table}[htbp]
\centering
\caption{\textbf{Notes written on posts where our LLM writer
abstained.} AI-written notes refer to notes produced by other AI writers.
Percentages are calculated separately within each abstention category.}
\label{tab:abstention_followup}
\small
\begin{tabular}{lcccc}
\toprule
& Human note & Human CRH note & Other AI note & Other AI CRH note \\
\midrule
No note needed
& 38.55\% & 7.76\% & 9.98\% & 1.99\% \\

Insufficient evidence
& 45.09\% & 12.22\% & 11.22\% & 2.20\% \\
\bottomrule
\end{tabular}
\end{table}

\begin{figure}[H]
  \centering
  \includegraphics[width=\textwidth]
  {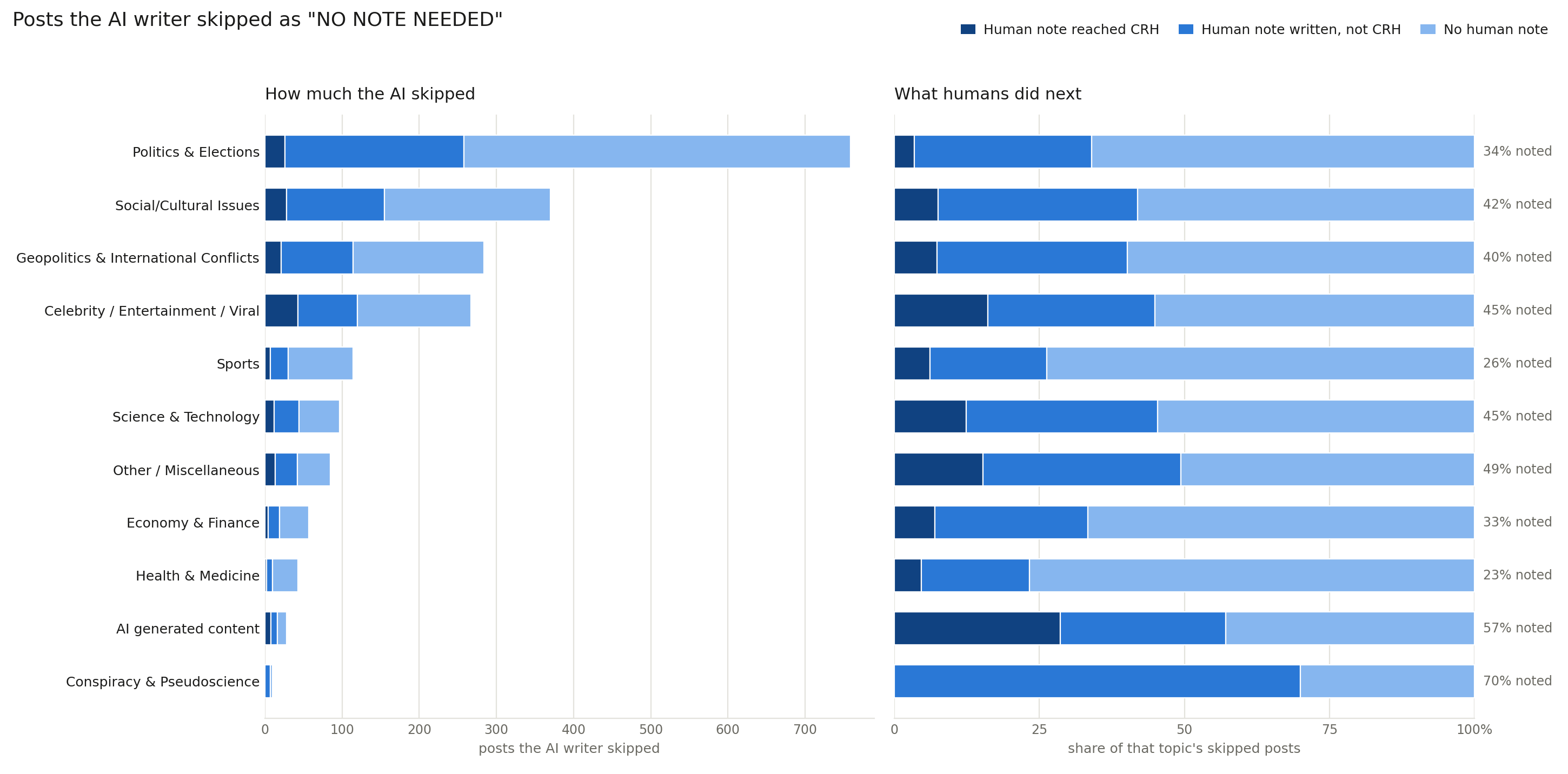}
  \caption{\textbf{Topic-specific outcomes for posts classified as not
  needing a note.} The left panel reports raw count of posts. The right panel reports the corresponding proportions within each topic. Percentages printed to the right indicate the share receiving any human-written note.}
  \label{fig:abstain_no_note_needed}
\end{figure}

\begin{figure}[H]
  \centering
  \includegraphics[width=\textwidth]
  {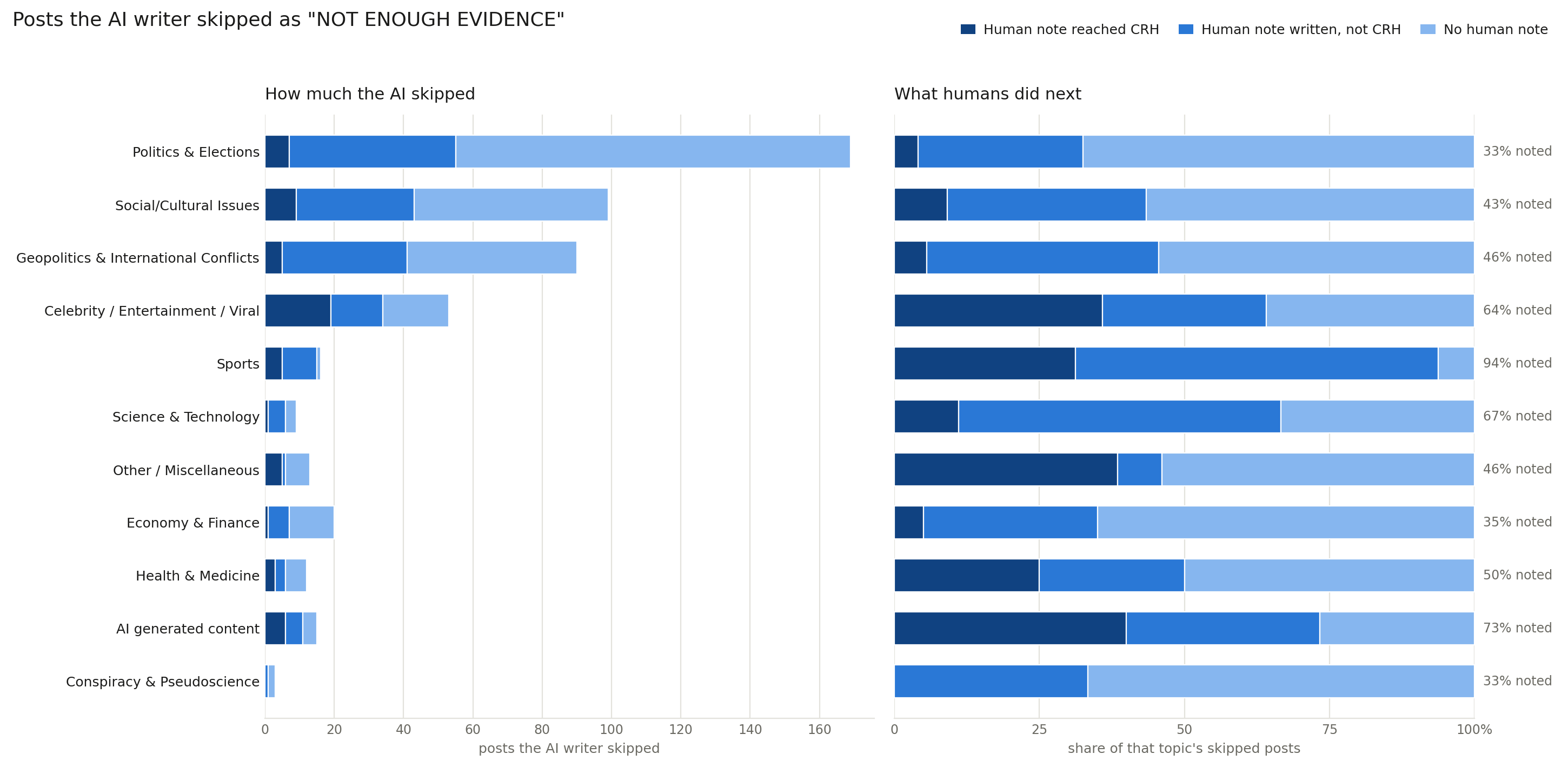}
  \caption{\textbf{Topic-specific outcomes for posts skipped because of
  insufficient evidence.} The left panel reports raw count of posts. The right panel reports the corresponding proportions within each topic. Percentages printed to the right indicate the share receiving any human-written note.}
  \label{fig:abstain_insufficient_evidence}
\end{figure}

Figures~\ref{fig:abstain_no_note_needed}
and~\ref{fig:abstain_insufficient_evidence} further describe human writing
outcomes by topic. The left panels show the number of posts on which the
writer abstained, while the right panels show the proportion that received
no human note, at least one human note but none reached CRH, or at least one human CRH note.

Posts classified as not needing a note were concentrated in politics and
elections, social and cultural issues, geopolitics and international
conflicts, and celebrity, entertainment, or viral content. Human contributors
subsequently annotated fewer than half of the skipped posts in most topics.

The insufficient-evidence category was smaller and showed a different
topic-specific pattern. Human contributors subsequently annotated 94\% of
sports posts and 73\% of AI-generated-content posts. By contrast, only 33\% of politics and elections posts. These patterns suggest that insufficient evidence sometimes reflects limitations in the writer's available evidence or tools rather than an absence of note-worthy content.

\section{LLM Writing Pipeline Prompts}
\label{app:prompts}

\subsection{Evidence Retrieval Prompt}
\label{app:prompt_research}

\begin{lstlisting}
"""Investigate whether the X post below is misleading. Images or video previews
from the post are provided if they exist.

Step 1: Understand the post and its context
- Find the post by post id {post_id} on X. If not found, search the post's main
  text on X.
- Identify the post author. Also note any signals that the account is
  parody/satire (e.g. bio, handle, verification).
- Fetch the thread and top comments. Understand how others interpret the post
  (e.g. joke/satire, potentially misleading, expressing an opinion).
- Summarize the post context in 1-2 sentences without rewording the post itself.
  Include who the OP is and how others interpret the post, especially if comments
  suggest it's a joke or satire, or provide potential fact-check directions.
  For example: "The OP is Sen. X. Many comments say his claim about X is
  unverified." or "The OP is a parody account, and commenters are laughing and
  treating it as a joke."

Step 2: Search for evidence
1. Search both the web and X for factual sources that refute or confirm the
   post's claims. Use the post context to guide your search if it could provide
   potential fact-check directions.
2. Aim for {target_url_count} pieces of evidence / URLs if possible.
3. For each source, include the URL and a brief note describing how it verifies
   or challenges the post. Include the publication date of the source if
   available.
4. Cover outlets across the ideological spectrum (left, center, right).
   Overlapping reasoning is acceptable when it comes from different publishers.
5. Prioritize evidence that is relevant, solid, and up to date.

Target post (ID: {post_id}):
{post}

Your response should be returned as a JSON object with the following structure:
```
{{
   "post_context": "one/two-sentence summary of the post context",
   "research": [
       {{"url": "url1", "description": "how the content of the URL fact-checks the post"}},
       ...
   ]
}}
```
If you cannot find sufficient evidence to fact-check the post, return an empty
research array."""
\end{lstlisting}

\subsection{Note Triage Prompt}
\label{app:prompt_decision}

\noindent\textbf{System prompt:}

\begin{lstlisting}
"""Decide whether a post needs a Community Note based on the provided evidence.

Returns:
    - "WRITE NOTE" if a note should be written
    - "NO NOTE NEEDED" if the post doesn't need a note
    - "NOT ENOUGH EVIDENCE" if there's insufficient evidence
"""
\end{lstlisting}

\noindent\textbf{User prompt:}

\begin{lstlisting}
"""You are a Community Notes writer. Your job is to decide if the target post
could be perceived as misleading and whether it needs a community note to address
its issues. The output should be one of: "WRITE NOTE", "NO NOTE NEEDED", or
"NOT ENOUGH EVIDENCE".

Task rules
- Focus on the main claims of the post, not trivial errors.
- If media is included, use any legible text, recognizable logos/landmarks, and
  clearly identifiable public figures as part of the claim.
- If the target post is a quote/reply, evaluate the target post, using the
  quoted/replied post only as context.
- Use the context information (author info and audience reactions) to inform your
  decision. They are helpful for understanding whether the post is likely to be
  perceived as misleading.
- If you are unsure, err on the side of NO NOTE NEEDED.

Decision logic
Output NO NOTE NEEDED if:
- The OP is a satire/parody account, or the post is joking/sarcastic/exaggerated
  to be ironic, and commenters are interpreting it that way (with no strong signs
  of misunderstanding); or
- The post is mostly opinion, subjective takes, or personal experience; or
- The post contains no major factual claims, or the claims are not verifiable; or
- The post contains factual claims, but the provided evidence indicates those
  claims are accurate or not meaningfully misleading.

If the post contains major factual, verifiable claims and the evidence is
relevant:
- Output NOT ENOUGH EVIDENCE if the evidence is weak, mixed, or insufficient to
  confidently verify or refute the main claims. When unsure about the
  strength/sufficiency of the evidence, err on NOT ENOUGH EVIDENCE.
- Output WRITE NOTE only if the evidence clearly shows that the post's main
  claims are false or misleading in a way that could misinform a reasonable
  reader.

Post:
{post}

Additional context information about the post:
{post_context}

Evidence:
{evidence}

Output only one of: "WRITE NOTE", "NO NOTE NEEDED", or "NOT ENOUGH EVIDENCE".
Err on the side of NO NOTE NEEDED if unsure."""
\end{lstlisting}

\subsection{Note Generation Prompt}
\label{app:prompt_write}

\noindent\textbf{System prompt:}

\begin{lstlisting}
"""You are a helpful fact-checking assistant.
Your goal is to write good Community Notes that would be approved helpful by
people with different viewpoints.
Do not invent facts or make claims that are not supported by the provided
evidence."""
\end{lstlisting}

\noindent\textbf{User prompt:}

\begin{lstlisting}
"""Task: Write a community note for the target post below. Images or video
previews from the target post may be provided; if present, analyze any legible
text (OCR), recognizable logos/landmarks, and confidently identifiable public
figures to inform the note. Additional context provides post author details and
audience reactions. If the target post quotes/replies to another post, use it
only for context and focus on the target post.

Hard Constraints:
1. The note is written to explain why the post is misleading and add additional
   context to the post. Focus on primary claim(s) of the post rather than
   trivial details.
2. The note must be grounded in the provided evidence and should cite the URL of
   the evidence it uses. At least one URL must be cited.
3. Keep the note strictly under 280 characters. Stay neutral and clear.
4. No hashtags, emojis, unnecessary words. No markdown, brackets, or parentheses
   around URLs. Do not mention "this note" or "the prompt."

Target post:
```
{post}
```

Additional context about the post:
```
{post_context}
```

Allowed evidence sources:
```
{evidence}
```

Output only the final note (at most 280 characters)."""
\end{lstlisting}

\subsection{Topic Classification Prompt}
\label{app:prompt_topic}

\begin{lstlisting}
"""Classify the following X/Twitter post into exactly one of these categories:

- Politics & Elections: U.S./global elections, politicians, voting, partisan
  claims, policy debates.
- Geopolitics & International Conflicts: Wars, foreign policy, terrorism,
  diplomacy, country-specific events (e.g., Ukraine, Israel-Palestine,
  China-Taiwan).
- Health & Medicine: Diseases, treatments, vaccines, public health policies,
  medical advice, COVID.
- Social/Cultural Issues: Abortion, gender/LGBTQ+, race/DEI, guns,
  crime/justice, education, religion.
- Economy & Finance: Inflation, jobs, taxes, crypto, markets, inequality.
- Science & Technology: Climate change, AI, space, gadgets (non-health tech).
- Conspiracy & Pseudoscience: General conspiracies, election fraud claims not
  tied to active politics, QAnon-style, flat earth, etc.
- Celebrity / Entertainment / Viral: Non-political hoaxes, celebrity drama.
- Sports: Sports events, sports scandals, and relevant discussion.
- AI generated content: AI generated / modified content.
- Other / Miscellaneous: Everything else (weather, personal, ads, neutral news
  without controversy).

Post text:
{tweet_text}
{context_section}
Answer with only the exact category name from the list above
(e.g., "Politics & Elections" or "Other / Miscellaneous")."""
\end{lstlisting}

\end{document}